\documentclass[iop]{emulateapj}

\usepackage{graphicx}
\usepackage{amsmath}
\usepackage{subfigure}
\usepackage{float}
\usepackage{natbib} 
\usepackage{hyperref}
\usepackage{tabularx}

\begin{document}

\title{Rotation periods of TESS Objects of Interest from the Magellan-TESS Survey with multiband photometry from Evryscope and TESS}

\author{Ward S. Howard\altaffilmark{1}, Johanna Teske\altaffilmark{2,**}, Hank Corbett\altaffilmark{1}, Nicholas M. Law\altaffilmark{1}, Sharon Xuesong Wang\altaffilmark{3,4}, Jeffrey K. Ratzloff\altaffilmark{1}, Nathan W. Galliher\altaffilmark{1}, Ramses Gonzalez\altaffilmark{1}, Alan Vasquez Soto\altaffilmark{1}, Amy L. Glazier\altaffilmark{1}, Joshua Haislip\altaffilmark{1}}

\altaffiltext{1}{Department of Physics and Astronomy, University of North Carolina at Chapel Hill, Chapel Hill, NC 27599-3255, USA}
\altaffiltext{2}{Earth and Planets Laboratory, Carnegie Institution for Science, 5241 Broad Branch Road, NW, Washington, DC 20015, USA}
\altaffiltext{**}{Some of this work was completed while this author was \\ a NASA Hubble Fellow at the Observatories of the Carnegie \\ Institution for Science.}
\altaffiltext{3}{Department of Astronomy, Tsinghua University, Beijing 100084, China}
\altaffiltext{4}{Observatories of the Carnegie Institution for Science, 813 Santa Barbara Street, Pasadena, CA 91101}

\email[$\star$~E-mail:~]{wshoward@unc.edu}

\begin{abstract}
Stellar RV jitter due to surface activity may bias the RV semi-amplitude and mass of rocky planets. The amplitude of the jitter may be estimated from the uncertainty in the rotation period, allowing the mass to be more accurately obtained. We find candidate rotation periods for 17 out of 35 TESS Objects of Interest (TOI) hosting $<$3 R$_\oplus$ planets as part of the Magellan-TESS Survey, which is the first-ever statistically robust study of exoplanet masses and radii across the photo-evaporation gap. Seven periods are $\geq$3$\sigma$ detections, two are $\geq$1.5$\sigma$, and 8 show plausible variability but the periods remain unconfirmed. The other 18 TOIs are non-detections. Candidate rotators include the host stars of the confirmed planets L 168-9 b, the HD 21749 system, LTT 1445 A b, TOI 1062 b, and the L 98-59 system. 13 candidates have no counterpart in the 1000 TOI rotation catalog of Canto Martins et al. (2020). We find periods for G3-M3 dwarfs using combined light curves from TESS and the Evryscope all-sky array of small telescopes, sometimes with longer periods than would be possible with TESS alone. Secure periods range from 1.4 to 26 d with Evryscope-measured photometric amplitudes as small as 2.1 mmag in $g^{\prime}$. We also apply Monte Carlo sampling and a Gaussian Process stellar activity model from \texttt{exoplanet} to the TESS light curves of 6 TOIs to confirm the Evryscope periods.
\end{abstract}


\maketitle

\section{Introduction}

A dichotomy in the radii of small ($<$4 R$_{\oplus}$) exoplanets has been confirmed in numerous studies, e.g. \citealt{Fulton2017, Fulton_Petigura2018, Van_Eylen2018, Martinez2019, MacDonald2019}. A ``radius gap" in the relative occurrence rates of small planets appears at $\sim$1.8 R$_\oplus$ \citep{Fulton2017}. It is likely the gap is explained by two populations of planets: one with a significant H/He envelope around the rocky core and another without an envelope. Planets without an envelope formed in conditions preventing primordial envelope development, or lost their envelope \citep{Fulton_Petigura2018}. Several mechanisms responsible for the mass loss driving the observed radius gap have been proposed, including photo-evaporation and core-powered mass loss. Photo-evaporation primarily occurs in young planetary systems as X-ray and extreme UV emission from the host star efficiently removes volatiles from the planetary atmosphere \citep{Owen_Jackson2012, Lopez2012}, while core-powered mass loss occurs over $\sim$1 Gyr timescales and is due to Parker wind escape driven by primordial heat from the core \citep{Ginzburg2016, Ginzburg2018}. Inward drift of the radius gap at lower incident fluxes has been interpreted as evidence favoring photo-evaporation \citep{Van_Eylen2018, Carrera2018}. \citet{Loyd2020} find neither mechanism is strongly favored by current statistics, but a 2$\times$ increase in the population of precisely-characterized small planets may be able to remove the ambiguity. In addition to post-formation processes, it is possible planets without H/He envelopes formed in-situ \citep{Hansen_Murray2013, Chiang_Laughlin2013, Lee2014, Lee_Chiang2015, Lee_Chiang2016}; planets with H/He envelopes may have formed in different environments at larger orbits and have since migrated inwards \citep{Cossou2014, Raymond_Cossou2014, Schlichting2014}. 

A key step in distinguishing between the physical mechanisms responsible for the radius gap is the unbiased measurement of many exoplanet masses and radii (and therefore densities) across the radius gap \citep{Loyd2020}. If planet densities do not correlate with incident stellar fluxes, then processes beyond photo-evaporation are at work. Precision measurements of exoplanet masses are more challenging in the \textit{Kepler} sample where the radius gap has been most clearly observed because the host stars are often faint \footnote{https://exoplanetarchive.ipac.caltech.edu}. The population of nearby planets with masses suffer from statistical biases: the masses of small planets are generally published only for those planets where the RV semi-amplitude is much larger than the noise, leading to artificially high mass estimates at a given radius. Furthermore, population studies of exoplanet density do not robustly account for selection biases in RV follow-up \citep{Montet2018, Burt2018}.

The Magellan-TESS survey (MTS) is designed to account for selection biases in masses and in the RV follow-up target selection. The MTS is performing dedicated RV follow-up of dozens of 1-3 R$_\oplus$ transiting planets detected by TESS around nearby stars bright enough for RV follow-up \citep{Teske2020}. The narrow 1-3 R$_\oplus$ range is selected to provide as many targets as possible near the gap. The MTS is the first statistically robust survey of exoplanet densities. All mass constraints will be published to prevent biased mass estimates, not just those planets with semi-amplitudes 6$\sigma$ above the noise. Furthermore, all MTS targets are chosen on the basis of a simple and reproducible selection function. The selection function was chosen and then fixed prior to the start of any RV observations, enabling the true population of exoplanet densities to be backed out of the observed sample.  

Stellar activity is the dominant source of noise in RV observations of small planets \citep{robertson2014}. The rotation of starspots induces correlated noise in RV measurements. These spots may also brighten or dim over several rotation periods, further altering the RV signal \citep{Giles2017}. Stellar activity signals may change the RV semi-amplitude of the planet \citep{Haywood2018, Damasso2019}, or even result in false-positive detections of exoplanets \citep{robertson2014, Robertson2015}. Rotation-induced variability may be used to measure the stellar rotation period, $P_\mathrm{Rot}$. 

We measure $P_\mathrm{Rot}$ and its associated uncertainty to provide an input for later estimation of the amplitude of the stellar RV jitter resulting from the rotational variability. The MTS selection function prioritizes targets with smaller jitter values to obtain precise measurements of the RV semi-amplitudes of small planets. Careful planning in the cadence of RV observations allows the detection of planetary signals smaller than the activity signals: if the rotation period is known, coherent activity-induced variation in RVs observed within each cycle may be clearly identified and removed \citep{LopezMorales2016, Haywood2018}. A global fit to the RV time series that includes both the planet signal and the stellar rotation period may provide increased accuracy when measuring the RV semi-amplitude and mass, e.g. \citep{Haywood2014,Rajpaul2015,LopezMorales2016,Kosiarek2019,Kosiarek2019b}. For example, \citet{Kosiarek2019,Kosiarek2019b} isolate and remove stellar rotation signals from planetary mass signals using a Gaussian Process likelihood model with terms for both the planet and star's RV modulation. For MTS targets that are selected for RV follow-up, we use $P_\mathrm{Rot}$ to inform the priors in a Gaussian Process (GP) Keplerian fit to properly account for the stellar jitter signals in the RVs.

Precise measurement of the rotation periods of stars hosting TESS planet candidates, or TESS Objects of Interest (TOIs), is difficult for periods longer than $\sim$14 d using 28 d TESS light curves alone \citep{VanderPlas2018}. However, longer periods have been obtained from multi-sector light curves \citep{Martins2020}. \citet{Martins2020} has characterized the rotation periods of hundreds of TOIs in the TESS light curves alone. Confidently-detected periods extend out to $\sim$13 d, with longer-period detections becoming both less frequent and more dubious.

Long-term ground-based photometry has been shown to be effective at recovering the small-amplitude signals of rotators, and at periods from 10$^{-1}$-10$^2$ d, e.g. \citealt{Newton2016,Oelkers2018,Newton2018, Howard2019b}. To date, ground-based photometry from the Kilodegree Extremely Little Telescope (KELT; \citealt{Pepper2004}), the Wide Angle Search for Planets (WASP;\citealt{Pollacco2006}), MEarth \citet{Nutzman2008}, the All Sky Automated Survey (ASAS; \citealt{Pojmanski1997}), and other surveys has been used to constrain the rotation periods of a number of TOIs, e.g. \citealt{Benatti2019,Dragomir2019,Crossfield2019,Astudillo-Defru2020,Shporer2020}. The rotation period of TOI 200 (DS Tuc Ab) was verified in both the TESS and ASAS light curve data by \citet{Benatti2019}; determining periods with a combination of TESS and ground-based data has several advantages over using only one survey. A combination of TESS and ground based monitoring to identify and vet P$_{rot}>28$ d rotation signals removes systematic periodicity in each survey. A ground-based and spaced-based survey are likely to exhibit different systematics, allowing us to leverage each survey against the systematics of the other. This process effectively increases the sensitivity of the ground-based survey to small-amplitude rotators at periods that are longer than the observations spanned by a single TESS sector; if a 28 d TESS light curve contains an incomplete rotation, only the periodogram peaks at longer periods need be examined in the ground-based data. This prior on the period search range decreases the noise floor of the periodogram. Long-term ground-based monitoring also decreases the period uncertainty \citep{VanderPlas2018} and captures evidence of differential rotation and spot evolution via periodogram stacking of different seasons \citep{Haywood2018, Kosiarek2019}.

The Evryscope \citep{Evryscope2015,Ratzloff2019} observes all bright ($g^{\prime}<$15) stars in the South. The Evryscope is an array of small telescopes simultaneously imaging the entire accessible sky. Evryscope light curves allow detection of significantly longer rotation periods than from TESS data alone: while TESS observes each star for $\sim$28 days in the red at high photometric precision (and twice this time span in the Extended Mission), Evryscope observes each star for 2+ years in the blue at moderate precision. We combine Evryscope and TESS photometry to measure or constrain the rotation periods for 35 TOIs as part of the MTS.

In Section \ref{EvryFlare} of this work, we describe the Evryscope, light curve generation, and rotation period observations. We also describe the TESS observations. In Section \ref{evr_p_rot_measurements}, we describe rotation period detection in Evryscope and TESS and estimation of period uncertainties. In Section \ref{period_FAP}, we describe our objective criteria for assessing Evryscope+TESS periodograms. In Section \ref{results}, we describe rotation period detections and non-detections in TOIs highly ranked by the MTS metric and therefore candidates for mass measurement. In Section \ref{GP_section} we compare our rotation periods against Gaussian Process stellar rotation models with the TESS light curves. In Section \ref{discuss_conclude}, we summarize our results and conclude.

\section{Photometry}\label{EvryFlare}
We discover rotation periods in photometry from the TESS and Evryscope surveys.

\subsection{Evryscope observations}\label{evryscope_observations}
As part of the Evryscope survey of all bright Southern stars, we discover many variable stars, including stellar rotators. Evryscope-South is located at Cerro Tololo Inter-American Observatory in Chile, and Evryscope-North is located at Mount Laguna Observatory in California, USA. Each Evryscope unit is an all sky array of small telescopes with an instantaneous footprint of 8150 square degrees, covering a total of 18,400 square degrees as the Earth rotates each night. Evryscope-South is optimized for high cadence photometry of bright, nearby stars, with a two-minute cadence in \textit{g}\textsuperscript{$\prime$}~\citep{Evryscope2015} and a typical dark-sky limiting magnitude of \textit{g}\textsuperscript{$\prime$}=16. Each night, Evryscope performs continuous monitoring of the accessible sky down to an airmass of two and at a resolution of 13\arcsec pixel$^{-1}$ for $\sim$6 hours. The system accomplishes this coverage by employing a ``ratchet" strategy that tracks the sky for 2 hours before ratcheting back into the initial position and continuing observations \citep{Ratzloff2019}.

Evryscope-South has taken 3.0 million raw images, which are stored as $\sim$250~TB of data. Evryscope images are processed in real-time with a custom data reduction pipeline \citep{Law2016,Ratzloff2019}. Each 28.8 MPix Evryscope image is calibrated using a custom wide-field astrometric solution algorithm. Background modeling and subtraction are carefully performed before raw photometry is extracted within forced-apertures at coordinates in an Evryscope catalog of 3M known source positions. This catalog includes all stars brighter than \textit{g}\textsuperscript{$\prime$}=15, fainter cool stars, white dwarfs, and a number of other types of targets. Light curves are then generated across the Southern sky by differential photometry in small regions on the sky with carefully-selected reference stars and across several apertures \citep{Ratzloff2019}. Two iterations of the SysRem detrending algorithm remove most large systematics \citep{tamuz2005}. For reference, we note an Evryscope $g^{\prime}$ magnitude of 9 approximately corresponds to a TESS magnitude of 7, and a $g^{\prime}$ magnitude of 15 approximately corresponds to a TESS magnitude of 13.

We periodically regenerate the entire database of Evryscope light curves in order to incorporate recent observations and to improve the photometric precision. At the time the data was analyzed for the present work, the Evryscope light curve database spanned two years of observations, averaging 32,000 epochs per star (with factors of several increases to this number closer to the South Celestial Pole). Depending upon the level of stellar crowding, light curves of bright stars (\textit{g}\textsuperscript{$\prime$}=10) reach 6 mmag to 1\% photometric precision. Evryscope light curves of dim stars (\textit{g}\textsuperscript{$\prime$}=15) reach comparable precision to TESS, attaining 10\% photometric precision \citep{Ratzloff2019}. In between light curve database updates, we may query light curves of individual sources not in the standard database at high computational cost using a separate Evryscope pipeline, Evryscope Fast Transient Engine (EFTE; \citealt{Corbett2020}). The photometric performance and stability of the EFTE light curves is comparable to light curves from the standard pipeline. More details on the EFTE pipeline are found in \citet{Corbett2020}. We use light curves from both the standard and EFTE pipelines, which are tracked in the machine-readable version of Table \ref{table:rotation_per_tab}.

\subsection{TESS observations}\label{tess_observations}
The Transiting Exoplanet Survey Satellite (TESS; \citealt{Ricker_TESS}) primary mission looked for transiting exoplanets across the entire sky, split into 26 sectors. TESS observed each sector continuously with four 10.5 cm optical telescopes in a red (600-1000 nm) bandpass for 28 days at 21$\arcsec$ pixel$^{-1}$. TESS is now operating in an extended mission, which will extend its observing baseline from 28 d to 56+  d for most of the sky. Calibrated, short-cadence TESS light curves of each TOI were downloaded from MAST\footnote{https://mast.stsci.edu}. We selected Simple Aperture Photometry (SAP) light curves rather than Pre-search Data Conditioning (PDC) ones to avoid removing real astrophysical variability. For the three MTS TOIs without 2 min cadence TESS SAP light curves, we construct a systematics-corrected light curve at 30 min cadence from the TESS full frame images (FFI) using the \texttt{eleanor} pipeline \citep{Feinstein2019}. We extract postage stamps of height=15, width=15, and a background size=31. We do not use the \texttt{eleanor} features for removing light curve systematics using either the point spread function or principal component analysis options.

\subsection{Characterizing the TOI sample}\label{stellar_astrophysics}
Each TOI in the MTS is selected for RV follow-up on the basis of a merit function defined in \citet{Teske2020}. The targets in this work are those that have passed our first selection using the merit function, but have not yet been down-selected according to their various activity levels. In this paper, we estimate the spectral type of each star using (in order of priority) confirmation papers for published TESS planets, then SIMBAD \citep{Wenger2000}, and lastly from the TESS Input Catalog (TIC; \citealt{Stassun2019}) and ExoFOP-TESS via effective temperatures and a temperature-to spectral-type conversion from \citet{Kraus2007}. The spectral types are tabulated in Table \ref{table:rotation_per_tab}. The TESS magnitude and stellar distance are obtained from the TIC and EXOFOP-TESS. The $g^{\prime}$ magnitude is obtained from the AAVSO Photometric All Sky Survey (APASS) DR9 \citep{Henden2016}.

\section{Rotation period discovery and characterization}\label{evr_p_rot_measurements}
We search for photometric rotation periods by computing the Lomb-Scargle (LS) periodogram \citep{Lomb1976,Scargle1982,VanderPlas2018} of each Evryscope and TESS light curve.

\begin{figure}
	\centering
	{
		\includegraphics[trim= 1 1 1 1,clip, width=3.4in]{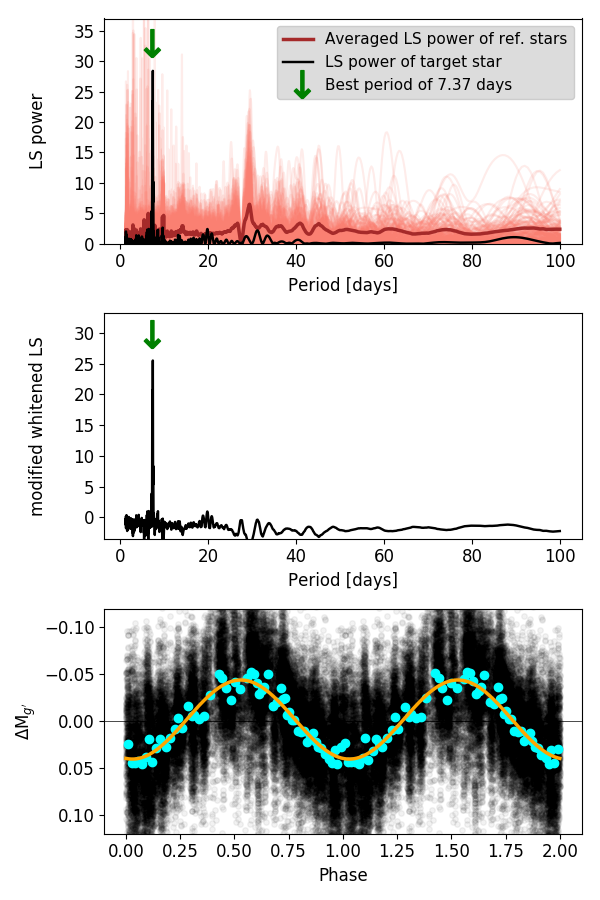}
	}
	\caption{Reproduced from \citet{Howard2019b}: An example photometric rotation period found in an Evryscope light curve. The LS periodograms of all stars are plotted on top of each other in a transparent red color, while the ``averaged" periodogram is plotted as a solid dark red line. The LS periodogram of the target star is plotted as a solid black line. The averaged LS periodogram is then subtracted from the LS periodogram of the target star and searched for the highest peak above the noise, as displayed in the middle panel of Figure \ref{fig:example_LS_detection}. The best period is denoted by a green arrow. In the bottom panel, we plot $\Delta$M$_{g^{\prime}}$ magnitude versus phase. A folded and binned Evryscope light curve is plotted in blue points and compared to the best-fit sine in orange.}
	\label{fig:example_LS_detection}
\end{figure}

\subsection{Simultaneous period detection in TESS and Evryscope}\label{simul_periods}
LS periodograms of TESS and Evryscope light curves complement each other. TESS light curves produce periodograms that are sensitive to low-amplitude variability due to rotation at short periods. Evryscope light curves produce periodograms sensitive to long-period rotators. Smaller-amplitude rotators may be identified in the Evryscope light curves if the period search range is constrained by prior information in the TESS periodogram. Furthermore, the FWHM of the LS peaks in Evryscope is at least an order of magnitude smaller than in TESS periodograms due to the longer baseline \citep{VanderPlas2018}. Each survey has unique systematic periodic structure, allowing each survey to vet periodic signals seen in the other one.

\subsubsection{TESS light curve and periodogram}\label{tess_periodogram_work}
We first inspect each TESS light curve by eye for any potential rotation. Signals may include a clear sinusoid, complex sinusoids, or an incomplete sinusoid. We use all available sectors of 2 minute cadence SAP flux light curves for each TOI. If none is available, we generate 30 minute cadence light curves from all available sectors of FFIs using \texttt{eleanor} as described in Section \ref{tess_observations}.

Systematics-affected epochs in each TESS light curve are identified by bad quality flags, rapid increases or decreases of flux common to multiple targets, or periods of unusually-high photometric scatter; these epochs are subsequently removed. If epochs are removed or pre-whitening is performed, ``yes" is indicated in the TESS whitening column of Table \ref{table:rotation_per_tab}. If short-period astrophysical variability in the light curve is impacted by systematics at longer periods\footnote{As an example of a long-period systematic, the SAP flux light curve of TOI 455 has a long-term linear trend superimposed on its rotation signal.}, we remove the longer period signals by subtracting a 1D Gaussian-blurred light curve with a blurring kernel approximately equal to the candidate rotation period. The candidate rotation period is identified in an initial visual assessment of each TESS light curve; kernel periods are given in the machine-readable version of Table \ref{table:rotation_per_tab}. The blurring kernel is defined by the 1$\sigma$ width in time of the Gaussians used to weight brightness values at those times. The choice of the kernel width determines the amount of smoothing applied to the light curve. We found in \citet{Howard2019b} that this blurring timescale is effective at removing periods nearly twice the blurring kernel.

The pre-whitening does however reduce the amplitude of the rotation-induced variability in the light curves of grade A and B (high-quality signal) rotators by up to $\sim$1 mmag. As a result, we are only sensitive to stellar rotation periods in the TESS light curves that have an amplitude of $\sim$1 mmag or greater. The Evryscope noise floor is $\sim$2 mmag, so smaller signals would not be confirm-able in both light curves. Most TESS-only signals would therefore be at best a dubious detection or more likely would be a result of systematic periodicity. We compute the LS periodogram of the final light curve for 20,000 uniform frequency steps over a test period range from 0.1 d out to the length of the light curve in d. We phase-fold the TESS light curve at the highest LS peaks to identify the best candidate periods.

\subsubsection{Evryscope light curve and periodogram}
Because the photometric scatter of Evryscope light curves is $\sim$1-10\%, we cannot identify candidate rotation periods in the unfolded Evryscope light curves. However, phase-folding the light curves of bright stars over 2+ years of data allows detection of rotators with amplitudes as low as 2-3 mmag.

We compute the LS periodogram of each light curve for 10,000 uniform frequency steps over a test period range of 1.25 to 100 d. We choose this period range because faster rotators are excluded due to the requirements of RV follow-up efforts. The MTS selected against candidates with very rapid rotation \citep{Teske2020}. However, our claim that periods faster than 1.25 d are not supported in our dataset is a result of how rapid rotation would either clearly imprint on the light curve or would result in no rotational modulation at all (a flat TESS light curve is not a candidate rotator as we describe in \S \ref{combined_approach}). The TESS light curves place very stringent constraints on the existence of fast rotation. The remaining periods in the TESS light curves have periods of 1.4 to $\sim$80 d. 

We subtract 27.5 day and 1 day best-fit sines from all light curves before computing the periodograms to suppress day-night and lunar cycles. To account for any resulting bias to $P_\mathrm{Rot}$ due to this procedure, we require A and B grade rotators to phase-fold to the same period as the phase-folded TESS light curve as described further in \S \ref{combined_approach}. LS power is computed as the LS periodogram peak of the target star over the noise of the target star periodogram. We define the noise of the periodogram as the standard deviation of periodogram power. We exclude a period region within 0.05 days of the detected peak from the noise computation. This helps to prevent the signal peak from biasing the noise measurement of the periodogram.

After the LS periodogram is computed for the target star as described in the previous paragraph, we next compute the modified pre-whitened Lomb-Scargle (MP-LS) periodogram as described in \citet{Howard2019b}. We briefly summarize this technique here. In order to correct for systematics-induced power during the period analysis, we compare the LS periodogram of the target star with the combined ensemble of LS periodograms of 284 other Evryscope light curves from stars in \citet{Howard2019}. Periodicity common to all light curves will increase the LS power of the target star at systematics-affected periods. We therefore compute the median and standard deviation of the detected LS powers of all stars at each test period from 1.25 to 100 d. We define the averaged LS periodogram as the 1$\sigma$ upper limit of the distribution of LS powers at each tested period. This process is illustrated in the top panel of Figure \ref{fig:example_LS_detection}, reproduced here from \citet{Howard2019b}. We subtract the averaged LS periodogram from the target star periodogram. This MP-LS periodogram allows the detection of high-amplitude astrophysical oscillations at periods that may also exhibit systematic periods. For such high-amplitude signals, the height of the peak is reduced in the MP-LS periodogram.

\subsection{Period detection in combined TESS and Evryscope data}\label{combined_approach}
We create a custom graphical user interface of figures containing three panels displaying the period information available from the Evryscope and TESS light curves (see Figure \ref{fig:TOI_134} for an example). In the top panel, we plot the TESS light curve of the TOI. In the second panel, we plot the TESS light curve and Evryscope light curve phase-folded to the candidate period. The phase-folded Evryscope light curve is binned. In the bottom panel, we plot the LS periodogram of the TESS light curve and the MP-LS periodogram of the Evryscope light curve.

We phase-fold the TESS and Evryscope light curves to each period where a high peak occurred in the periodograms. Periods are generally selected from greatest to least power until the most likely period identification is made according to the criteria for each confidence grade. The phase-folded Evryscope and TESS light curves are inspected by eye for a clear sinusoid at each candidate period (i.e. we look for the lowest-scatter-in-phase and simplest sinusoidal structure). This procedure involves choosing the highest LS peak evidenced in both surveys that minimizes the photometric scatter in the phase-folded light curves. Because both Evryscope and TESS light curves are converted to MJD as a common time zero-point prior to phase-folding, we also consider how well the phase-folded light curves align in their relative sinusoidal phases. The best candidate period selected from the LS peaks of each TOI is given a grade of ``A", ``B", ``U", or ``N." This nomenclature for grading the quality of rotation period candidates is adapted from \citet{Newton2016}; in our usage a grade of ``A" is considered a likely detection, ``B" is a possible detection, ``U" is highly dubious, and ``N" is no detection. 

When phase-folding the Evryscope light curve at a signal period, the light curve is first pre-whitened at periods significantly shorter and longer than the period of interest. This is done following the same method described in \S \ref{tess_periodogram_work} by subtracting a 1D Gaussian-blurred light curve with a blurring kernel equal to the period at which pre-whitening is desired. This process primarily removes noise associated with the day-night and other cycles that may obscure longer period trends, allowing periodicity such as that in Figure \ref{fig:first6_gradeA_TOI} to be readily observed in phase-folded light curves. These periods are given in the machine-readable version of Table \ref{table:rotation_per_tab}. We first verify we can recover grade A periods largely evident in TESS before assessing the more difficult grade B and U rotators. These cases are identified in Table \ref{table:rotation_per_tab}. For some rotators such as TOI 260, sinusoids are subtracted from the light curve at periods with strong systematics present. The removal of the sines can help decrease the scatter observed in the phase-folded and binned Evryscope light curves. Sines are not subtracted at periods that remove the target signal or that seem to create new periodicity that was not already observable in the phase folded light curves. The cases where sines were fit are given in \S \ref{results}.

The combination of Evryscope and TESS light curves also helps to minimize the effects of aliasing on our detections. For example, while Evryscope light curves might display annual aliases, TESS will have aliases of other signals. Suppose an Evryscope alias alters the LS power of the target signal. If an alias appears to be present, we phase-fold the other light curve from TESS at the original Evryscope peak and the second candidate peak to see which is the true peak.

\begin{turnpage}

\begin{table*}

\caption{Rotation Periods of 1-3 R$_\oplus$ TOIs Observed by Evryscope and TESS}
\begin{tabular}{p{0.9cm} p{0.9cm} p{1.1cm} p{1.1cm} p{1.1cm} p{1.1cm} p{1.5cm} p{1.0cm} p{1.3cm} p{1.1cm} p{1.1cm} p{1.1cm} p{1.1cm} p{1.1cm} p{1.2cm} p{1.2cm}}
\hline
 &  &  &  &  &  &  &  &  &  &  &  &  &  & \\
TOI & P$_\mathrm{Rot}$ & Err. in \newline P$_\mathrm{Rot}$ & Grade & FAP$_\mathrm{tot}$ & SpT & TESS lc & TESS \newline ampl. rot. & EVR \newline ampl. rot. & EVR obs. time & EVR num. epochs & EVR whiten? & TESS whiten? & GP P$_\mathrm{Rot}$ & GP err. in P$_\mathrm{Rot}$ \\
 & [d] & [d] &  & [\%] &  &  & [$\Delta$T] & [$\Delta g^{\prime}$] & [yr] &  &  &  & [d] & [d] \\

\hline
 &  &  &  &  &  &  &  &  &  &  &  &  &  & \\

134.01 & 30.7 & 0.9 & A & 11 & M1V & SAP\_FLUX & 0.00405 & 0.0071 & 1.95 & 32833 & yes & no & no & no \\
177.01 & 17.6 & 0.3 & A & 0.1 & M3V & SAP\_FLUX & 0.00715 & 0.00715 & 2.46 & 24569 & yes & no & 17.9 & 0.5 \\
179.01 & 8.64 & 0.04 & A & 0.1 & K2V & SAP\_FLUX & 0.0047 & 0.00485 & 1.96 & 40471 & yes & no & 8.76 & 0.1 \\
260.01 & 15.8 & 0.3 & A & 11 & M0V & SAP\_FLUX & 0.00285 & 0.00325 & 2.11 & 14982 & yes & no & 15.3 & 2.1 \\
455.01 & 1.397 & 0.002 & A & - & M3V & SAP\_FLUX & 0.00165 & 0.004 & 2.08 & 17051 & yes & yes & 1.4 & 0.004 \\
724.01 & 10.38 & 0.09 & A & 0.1 & G9V & SAP\_FLUX & 0.0055 & 0.00465 & 1.99 & 42084 & yes & yes & no & no \\
1062.01 & 26.0 & 0.4 & A & 0.1 & K0V & SAP\_FLUX & 0.00175 & 0.0021 & 1.99 & 63783 & yes & no & no & no \\
1063.01 & 7.9 & 0.1 & A & 0.1 & G8V & SAP\_FLUX & 0.0096 & 0.00885 & 1.62 & 47899 & no & yes & 7.88 & 0.04 \\
1097.01 & 5.11 & 0.05 & A & 0.1 & G3V & eleanor & 0.00535 & 0.0054 & 1.93 & 42700 & yes & no & 5.1 & 0.06 \\
1116.01 & 16.0 & 0.3 & A & 0.1 & K0V & SAP\_FLUX & 0.0036 & 0.00295 & 1.99 & 42084 & no & yes & 15.9 & 1.1 \\
175.01 & 39.6 & 2.2 & B & 22 & M3V & SAP\_FLUX & 0.0017 & 0.0049 & 1.98 & 30088 & yes & yes & no & no \\
175.02 & 39.6 & 2.2 & B & 22 & M3V & SAP\_FLUX & 0.0017 & 0.0049 & 1.98 & 30088 & yes & yes & no & no \\
186.01 & 33.9 & 5.0 & B & 33 & K4.5V & SAP\_FLUX & 0.003 & 0.0035 & 2.46 & 29059 & yes & yes & no & no \\
461.01 & 15.2 & 0.2 & B & 30 & K1/2V & SAP\_FLUX & 0.0049 & 0.0038 & 1.96 & 17541 & yes & yes & no & no \\
697.01 & 13.6 & 0.2 & B & 24 & K0 & SAP\_FLUX & 0.00105 & 0.0013 & 1.97 & 39767 & yes & yes & no & no \\
776.01 & 22.7 & 0.4 & B & 24 & M1V & SAP\_FLUX & 0.0059 & 0.0095 & 2.45 & 33693 & yes & no & no & no \\
776.02 & 22.7 & 0.4 & B & 24 & M1V & SAP\_FLUX & 0.0059 & 0.0095 & 2.45 & 33693 & yes & no & no & no \\
836.02 & 9.0 & 1.3 & B & 100 & K7V & SAP\_FLUX & 0.003 & 0.0023 & 2.38 & 23533 & yes & no & no & no \\
836.01 & 9.0 & 1.3 & B & 100 & K7V & SAP\_FLUX & 0.003 & 0.0023 & 2.38 & 23533 & yes & no & no & no \\
913.01 & 32.0 & 7.0 & B & 37 & K2V & SAP\_FLUX & 0.0034 & 0.0034 & 2.97 & 104018 & yes & no & no & no \\
214.01 & 28.8 & 0.6 & U & - & G9.5V & SAP\_FLUX & 0.0011 & 0.00155 & 1.99 & 35722 & yes & yes & no & no \\
431.02 & 14.7 & 0.1 & U & - & K3V & SAP\_FLUX & 0.00165 & 0.0013 & 2.25 & 18624 & yes & yes & no & no \\
1233.04 & 41.0 & 1.4 & U & - & G5V & SAP\_FLUX & 0.0018 & 0.0033 & 1.99 & 46815 & yes & no & no & no \\
562.01 & 5.8 & 0.1 & U & - & M2.5V & SAP\_FLUX & 0.0006 & no & 1.96 & 23622 & yes & yes & no & no \\
1233.03 & 41.3 & 1.4 & U & - & G5V & SAP\_FLUX & 0.0018 & 0.0033 & 1.99 & 46815 & yes & no & no & no \\
402.01 & 16.2 & 0.3 & U & - & K1V & SAP\_FLUX & no & no & 2.39 & 10750 & yes & yes & no & no \\
174.01 & 29.8 & 0.7 & U & - & K3.5V & SAP\_FLUX & no & no & 1.99 & 50922 & no & yes & no & no \\
784.01 & 3.4 & 0.5 & U & - & G5V & SAP\_FLUX & no & no & 2.45 & 56606 & yes & yes & no & no \\
402.02 & 16.2 & 0.3 & U & - & K1V & SAP\_FLUX & no & no & 2.39 & 10750 & yes & yes & no & no \\
733.01 & 3.3 & 1.0 & U & - & G5V & SAP\_FLUX & no & no & 2.45 & 36854 & no & yes & no & no \\
174.02 & 29.8 & 0.7 & U & - & K3.5V & SAP\_FLUX & no & no & 1.99 & 50922 & no & yes & no & no \\
895.01 & 6.3 & 1.0 & U & - & G0V & eleanor & no & no & 1.96 & 9335 & yes & no & no & no \\
719.01 & 5.9 & 1.5 & U & - & G5V & SAP\_FLUX & no & no & 1.99 & 46449 & yes & yes & no & no \\
174.03 & 29.8 & 0.7 & U & - & K3.5V & SAP\_FLUX & no & no & 1.99 & 50922 & no & yes & no & no \\
283.01 & 64.8 & 7.0 & U & - & K0V & SAP\_FLUX & 0.0029 & 0.0029 & 1.78 & 47537 & yes & yes & no & no \\
286.01 & 44.6 & 5.0 & U & - & K0V & SAP\_FLUX & 0.00185 & 0.00185 & 2.97 & 66544 & yes & yes & no & no \\
286.02 & 44.6 & 5.0 & U & - & K0V & SAP\_FLUX & 0.00185 & 0.00185 & 2.97 & 66544 & yes & yes & no & no \\
262.01 & no & no & N & - & K0V & SAP\_FLUX & no & no & 2.39 & 9973 & no & no & no & no \\
652.01 & no & no & N & - & G2V & SAP\_FLUX & no & no & 2.45 & 16869 & no & no & no & no \\
141.01 & no & no & N & - & G1V & SAP\_FLUX & no & no & 1.77 & 41241 & no & no & no & no \\
 &  &  &  &  &  &  &  &  & ... &  &  &  &  & \\
 &  &  &  &  &  &  &  &  &  &  &  &  &  & \\
\hline
\end{tabular}
\label{table:rotation_per_tab}
{\newline\newline \textbf{Notes.} Parameters of 43 1-3 R$_\oplus$ TESS planets orbiting 35 unique stellar hosts (1 planet per row). This is a subset of the full table- the remaining 3 grade N rotators are in the machine-readable version. Columns displayed here are: TOI number, stellar rotation period, uncertainty on the period, the quality grade, FAP$_\mathrm{tot}$, spectral type, the source of the TESS light curve (i.e. 2 min cadence SAP FLUX or 30 min cadence \texttt{eleanor} data product), whether a 2 min cadence Evryscope light curve exists, the TESS-measured sinusoidal amplitude of rotation in $\Delta T$ mag, the Evryscope-measured sinusoidal amplitude of rotation in $\Delta g^{\prime}$ mag, the duration of Evryscope observations, the number of Evryscope epochs, a note whether the Evryscope light curve has been pre-whitened, a note whether the TESS light curve has had likely systematics-affected epochs removed and/or been pre-whitened, the GP measured period, and the uncertainty on the GP measured period. The machine-readable table also includes FAP$_\mathrm{LS}$ and FAP$_\phi$ values and the kernel periods at which the Evryscope and TESS light curves of grade A and B targets were pre-whitened.}
\end{table*}

\end{turnpage}

The criteria for assigning a grade is as follows:
\begin{itemize}
    \item A: This grade is assigned for likely candidates that have LS peaks in both the Evryscope and TESS periodograms. They must also demonstrate rotational modulation in both the TESS and Evryscope phase-folded light curves. An exception is made for TOI 455 as $\sim$10 high-amplitude complete cycles are present, leaving no doubt about the signal's existence However, TOI 455 is a close triple star system \citep{Winters2019}, so the component the signal is detected from is not certain.
    \item B: This grade is assigned for possible candidates that meet all but one criteria for an A-grade rotator. It is also assigned if all criteria are met but if there is some uncertainty in multiple criteria. For example, a rotator with an Evryscope MP-LS peak that only roughly aligned with the TESS LS peak but demonstrated sinusoidal modulation in the light curve of the same phase and comparable amplitude as in TESS would receive this grade. Comparable amplitudes must be within a factor of $\sim$3$\times$ agreement as $T$ and $g^{\prime}$ band variability may differ. Grade B rotators must display rotational modulation at the same period in both Evryscope and TESS phase-folded light curves.
    \item U: These are highly dubious periods with uncertainty in at least three of the criteria listed for an A-grade rotator, or else two criteria are entirely absent. These should not be trusted unless confirmed by previous studies. The signals most likely to be real are given in Section \ref{ugrade_rotators}.
    \item N: If no likely period identification can be made, the star is assigned a grade of ``N" for ``none." These stars may be either low-activity or have symmetric spot patterns that do not induce periodic oscillations in a light curve.  
\end{itemize}
We caution that the grade system is qualitative and not quantitative. This is because these criteria are designed to be a detection tool and not an objective statistical confirmation tool. While the \textit{grade} system is qualitative, we also develop an objective, quantitative system of criteria. Objective criteria for the confirmation of grade A and B candidates are given in Section \ref{FAP_obj} and \ref{inj_rec_tests}.

We also caution that we assume statistically significant grade A and B signals that also pass our FAP tests (\S \ref{FAP_obj}) are indeed due to rotational modulation. However, it is possible in some cases that aperiodic yet coherent activity such as the emergence and disappearance of spot complexes could cause aligned peaks in both the TESS and Evryscope periodograms. This is unlikely for grade A and B rotators as the TESS light curves must show signals consistent with rotational modulation to receive that grade. The clear rotational modulation in the unfolded TESS light curve required for a grade of A or B also helps to protect against the situation where a strong but spurious LS peak in 1 dataset and any peak in a second dataset may create a significant but false signal \citep{Koen2020}. For further examples of what believable rotational modulation looks like in TESS light curves of TOIs, see \citet{Martins2020}.

The combined analysis of TESS and Evryscope light curves assumes that rotational variability will be both persistent and coherent over the full monitoring baseline. If these assumptions are not largely upheld, we would likely not be able to confirm a detection from that TOI. Furthermore, the criterion requiring Evryscope data to fold to a coherent shape at the same period as TESS in the phase-folded graphs places strong limits on non-periodic variability being the dominant source of agreement between the surveys. Otherwise, the agreement in phase would not be persistent across 2 yr of Evryscope observations.

\subsection{Multiple stars in the aperture}\label{multi_stars}
We note that multiple stars may sometimes occur in the same aperture due to the 13" pix$^{-1}$ and 21" pix$^{-1}$ pixel scales of Evryscope and TESS, respectively. Among our grade A and B rotators, 9 targets cross-match with multiple \textit{Gaia} DR2 sources within 42" \citep{Gaia2016, Gaia2018} and 5-6 magnitudes of the target source. To determine whether the sinusoidal variability we report is from the bright TOI or the fainter source, we solve for the percent variation that the faint source would have to display to account for the flux amplitude we observe. If the faint source would have to vary by more than 100\% to produce the observed variability of the Evryscope light curve, then the TOI is the likely source of the variability. The targets where some uncertainty remains in which star is variable after this vetting procedure include TOIs 177, 186, 455, 913, and 1116. The next brightest source near the Gaia $G$=10.6 TOI 177 is a $G$=14.5 star, which would have to vary by 24\%. Such high-amplitude variability is rare in Evryscope data and would place this fainter star into the most active regime observed in \citet{Howard2019b}. Similarly-large amplitudes would be required and therefore similar constraints apply to the faint sources near TOIs 913 and 1116. \citet{Winters2019} notes the rotation from TOI 455 is likely from the BC component (whereas the A component is the host of planet TOI 455.01), leaving only the source of the variability observed from TOI 186 indeterminate.

\subsection{Measurement of period uncertainty }\label{period_error_boostrap}
We estimate the error of each candidate period to lie within the FWHM of the Evryscope periodogram peak, adjusting the period error upwards to lie within the much larger TESS periodogram peak FWHM when the Evryscope LS peak is indeterminate.

\section{Objective criteria for identifying the dominant period in TESS+Evryscope periodograms}\label{period_FAP}
While candidates are discovered using a mix of visual analysis of the Evryscope and TESS light curves and periodograms, an objective approach is needed to confirm that signals are observed in both datasets. Even when a rotational signal is apparent in TESS, a weak LS peak in the Evryscope data may be due to chance or systematics.

\begin{figure*}
	\centering
	{
		\includegraphics[trim= 1 1 1 1,clip, width=0.85\textwidth]{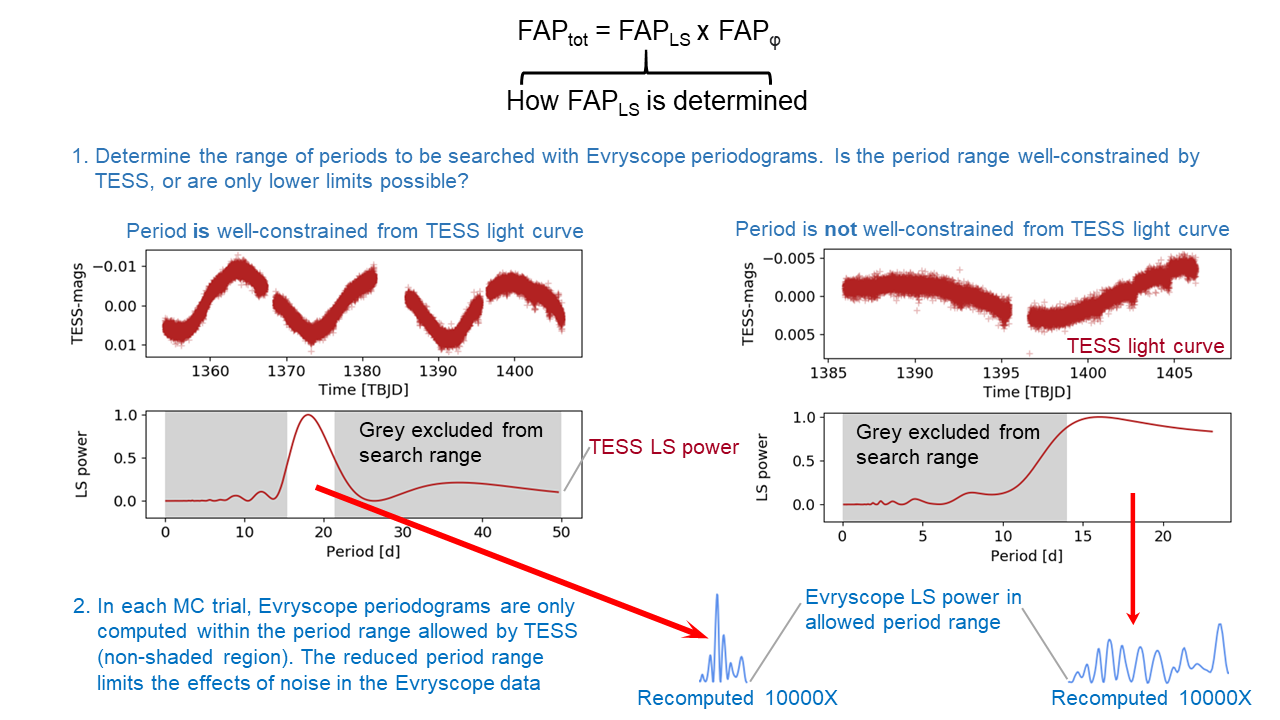}
	}
	\caption{How the FAP of the Evryscope LS periodogram is computed. First, the range of periods to be searched with the periodogram is determined using the TESS light curve (shown in red). When the period is well constrained by TESS (left panel), the period range to be searched in the Evryscope periodogram is given by the periods within the FWHM of the TESS LS periodogram (shown in red). When the period is not well constrained by TESS (right panel), only the lower limit on the possible periods may be placed as shown on the right. In both cases, Evryscope periodograms of random light curves with the same window function as the target star are computed 10000$\times$ and the FAP$_\mathrm{LS}$ is measured. Example periodograms in the TESS-constrained search range are shown in blue.}
	\label{fig:methods_peakLS}
\end{figure*}

\begin{figure*}
	\centering
	{
		\includegraphics[trim= 1 1 1 1,clip, width=0.85\textwidth]{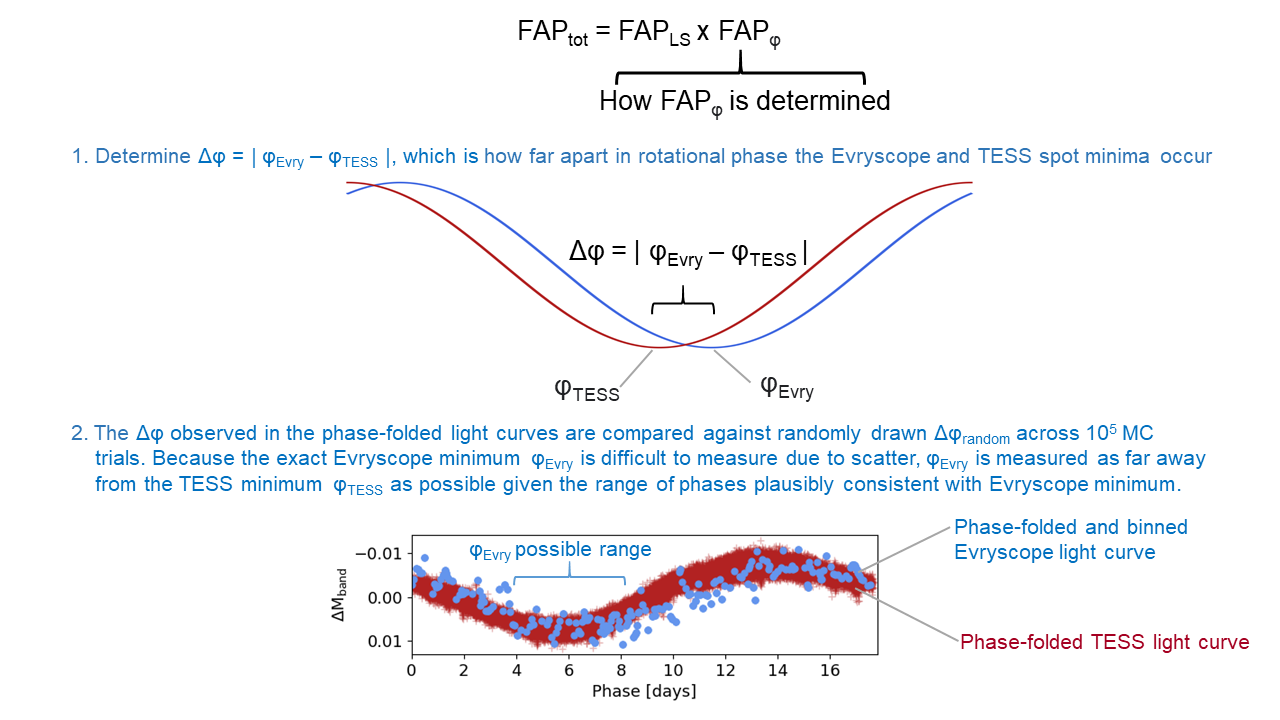}
	}
	\caption{How the FAP of the Evryscope versus TESS sinusoidal phase offset is computed. When the Evryscope (blue) and TESS (red) light curves are folded in phase to the detected period, the phases at which their sinusoidal troughs occur are compared. The sinusoidal troughs $\phi_\mathrm{Evry}$ and $\phi_\mathrm{TESS}$ occur at spot minima, when the dominant starspot is facing us and should therefore have similar values. The phase offset between the Evryscope and TESS trough phases $\Delta \phi = \phi_\mathrm{Evry} - \phi_\mathrm{TESS}$ therefore allows us to test if the agreement we observe is likely astrophysical or due to chance. 10$^5$ MC trials compare randomly-drawn phase offsets to the actual offset we observed and then the FAP$_\phi$ is computed.}
	\label{fig:methods_phaseLS}
\end{figure*}

\subsection{Evryscope+TESS false-alarm analysis}\label{FAP_obj}
MC tests are performed as described below to confirm the Evryscope+TESS detections with objective criteria:
\begin{enumerate}
    \item High precision TESS photometry is used to constrain the period search range for the false-alarm analysis of each candidate, as illustrated in Figure \ref{fig:methods_peakLS}. When several complete cycles of the rotation period are present in the TESS light curve, the minimum period in the periodogram search window $P_\mathrm{min}$ and the maximum period $P_\mathrm{max}$ are identified as the range of periods covered by the FWHM of the TESS LS signal as shown in the left panel of Figure \ref{fig:methods_peakLS}. When $\leq$1 periods are present in the TESS light curve, $P_\mathrm{min}$ is identified as the shortest-period sinusoid consistent with the observed peak-to-trough time in the TESS light curve as shown in the right panel of Figure \ref{fig:methods_peakLS}. $P_\mathrm{max}$ is set to 50 d, twice the longest-period secure detection in our dataset. This value is constant across every target in our sample for which $\leq$1 periods are present in the TESS light curve to avoid arbitrary selection effects in our objective criteria. The situation in which no periodicity is observed in TESS is not considered as this would result in a non-detection.
    \item We perform 10,000 Monte Carlo (MC) trials to determine how often peaks larger than the candidate signal occur from chance or systematics within the specified period range. In each trial, we shuffle the epochs and magnitudes in the light curve and then inject Evryscope systematics to prevent over-estimating the significance of the candidate signal. To inject systematics, an Evryscope light curve of another star without rotational modulation present is randomly selected and its magnitude values are added onto the randomly shuffled magnitudes. This process preserves both the window function and common systematics in the candidate light curve. Before the periodogram of the shuffled light curve is computed, it is pre-whitened with the exact same 1D Gaussian-blurred kernel that was applied to the candidate light curve in \S \ref{combined_approach}. 
    \item The LS (not MP-LS, which is defined in \S \ref{evr_p_rot_measurements}) periodogram is then computed and the highest peak in the range ($P_\mathrm{min}$, $P_\mathrm{max}$) is recorded. FAP$_\mathrm{LS}$ is defined as the fraction of MC trials with a peak higher than the candidate signal.
    \item In addition to the height of the Evryscope LS peak, agreement between the rotational phase of the Evryscope and TESS signals may be used to compute FAP$_\phi$, or the probability that the two signals have the same phase by chance. 10$^5$ MC trials are computed in which the offset in rotational phase between the Evryscope and TESS sinusoids $\Delta \phi = \phi_\mathrm{Evry} - \phi_\mathrm{TESS}$ is compared with randomly generated phase offsets $\Delta \phi_\mathrm{random}$. Phase offsets are measured in units of normalized phase. The shortest distance in phase may either pass through 1 and back through 0 or completely fall within the (0,1) range. The FAP is determined by dividing the number of MC trials in which $\Delta \phi_\mathrm{random} \leq \Delta \phi_\mathrm{actual} \leq 0.5$ by the total number of all trials. The process of computing phase offsets is illustrated in Figure \ref{fig:methods_phaseLS}. Because the Evryscope data is lower precision than TESS, $\phi_\mathrm{Evry}$ is sampled at both the most likely position of the Evryscope minimum/sinusoidal trough as well as at a value as far away from the TESS minimum/trough as is consistent with the phase-folded light curve. The largest offset from TESS in the allowed range of $\phi_\mathrm{Evry}$ values that gives the highest FAP$_\phi$ is used in the FAP$_\mathrm{tot}$ calculation, but both values (i.e. the most likely position of the Evryscope minimum, and the value as far away from the TESS minimum as allowed) are reported in the machine-readable version of Table \ref{table:rotation_per_tab}.
    \item The probability that an Evryscope signal is in fact the same as that seen in TESS depends on both the significance of the LS peak and also how well the phases agree. We define the total FAP of each candidate as FAP$_\mathrm{tot}$ = FAP$_\mathrm{LS}$ $\times$ FAP$_\phi$, assuming LS peak height and rotational phase are independent probabilities. 
    \item To ensure that multiplying the two probabilities  FAP$_\mathrm{LS}$ $\times$ FAP$_\phi$ accurately reflects a convolved distribution, we compare FAP$_\mathrm{tot}$ values against false alarm probabilities computed from a singular distribution, $\chi_\mathrm{dist}$(LS, $\Delta \phi$),
    \begin{equation}
    \chi_\mathrm{dist}(LS, \Delta \phi) = ||LS|| \cdot ||1/2- \Delta \phi||
    \label{eq:chi_rho_distrib}
    \end{equation}
    where LS are the randomized LS powers and $\Delta \phi$ are the randomized phase offset values. Each distribution is then compared with the observed $||$LS$_\mathrm{obs} || \cdot || 1/2-\Delta \phi_\mathrm{obs}||$ to generate a qualitative false alarm probability, FAP$_\chi$. We use 1/2-$\Delta \phi$ instead of $\Delta \phi$ to account for smaller phase offsets being stronger signals than large phase offsets, while higher LS powers are stronger signals. The LS and 1/2-$\Delta \phi_\mathrm{obs}$ distributions are normalized by their medians before multiplying to ensure two strong signals combine to make stronger signals instead of weaker ones. All of the 3$\sigma$ detections from FAP$_\mathrm{tot}$ are also 3$\sigma$ detections with FAP$_\chi$. Qualitative agreement at lower significance levels exists between FAP$_\chi$ values and FAP$_\mathrm{tot}$ values, although there are differences. For example, TOI 134 has a FAP$_\mathrm{tot}$ of 11\% but a FAP$_\chi$ of 33\%. TOI 175 has a FAP$_\mathrm{tot}$ of 22\% but a FAP$_\chi$ of 37\%. One of the most significant differences is for TOI 260, which has a FAP$_\mathrm{tot}$ of 10\% but a FAP$_\chi$ of 64\%. We also mention TOI 776 is a plausible non-detection with a FAP$_\mathrm{tot}$ of 24\%, but its FAP$_\chi$ is 100\%. Most FAP$_\chi$ are almost certainly overestimated. Since the conservative FAP$_\phi$ of TOI 260 is 11\% and is computed from a single $\Delta \phi$ distribution, adding even a 100\% FAP$_\mathrm{LS}$ should not invalidate the 11\% phase agreement. Since LS and $\Delta \phi$ should be independent, FAP determined from $\chi_\mathrm{dist}$ are much more dependent on the choice of normalization than FAP$_\mathrm{tot}$. FAP$_\chi$ can also significantly overestimate the true false-alarm probability due to the subtraction and normalization step. If the FAP determined exclusively from the Gaussian distribution of LS values is 10\%, a higher FAP$_\chi$ is probably unphysical. Likewise, if the chance of a random phase offset as good as the observed one is 10\%, a much higher FAP$_\chi$ is probably unphysical. We therefore adopt FAP$_\mathrm{tot}$ instead of FAP$_\chi$, using the latter for confirmation and illustration purposes only. 
\end{enumerate}
The FAP$_\mathrm{tot}$ of each TOI is given in Table \ref{table:rotation_per_tab}. The FAP$_\mathrm{LS}$ value, the best FAP$_\phi$ value, and the conservative FAP$_\phi$ value used for constructing FAP$_\mathrm{tot}$ are also given in the machine-readable version of Table \ref{table:rotation_per_tab}. It is important to note FAP$_\mathrm{tot}$ is conservatively constructed to give an upper limit. For example, a low SNR Evryscope signal may be astrophysical and apparent to the eye while not reaching a formal 3$\sigma$ detection threshold. This situation is likely for TOI 186 where the rotational period has previously been determined (e.g. \citealt{Gan2021}) and only one period must be tested in FAP tests. Since we do not employ all available information in our formalized false-alarm analysis, we do not claim TOI 186 as a secure detection from the FAP$_\mathrm{tot}$ formalism. We remind the reader this formalism must apply in the same way to all targets in the sample and additional information available for targets like TOI 186 therefore cannot be included. For example, additional information in this case includes a known $P_\mathrm{rot}$ and a best-value of the phase agreement in Table \ref{table:rotation_per_tab} much better than the conservative limit from that table.

Before measuring the FAP$_\mathrm{LS}$ value, the LS strength of the actual peak discovered from the Evryscope light curve must be ascertained. The actual LS signal from the Evryscope periodogram of each TOI is computed as the highest peak within the period uncertainty given in Table \ref{table:rotation_per_tab}. The highest LS power within the Evryscope peak is selected rather than the LS power at the exact $P_\mathrm{Rot}$ value from Table \ref{table:rotation_per_tab} to allow for slight differences between the preferred periods in Evryscope and TESS light curves (both of which informed the tabulated periods). A result of this process is that the FAP$_\mathrm{LS}$ values and the period error bars interact. When the period uncertainties are very large, nearby LS peaks due to random or correlated noise can upwardly bias the power. In some low-SNR cases such as TOI 186 these peaks are probably both part of the same signal, but modulated by power from the window function and correlated noise. For error bars greater than 2 d, these peak LS values are inspected by eye. In these cases, if the highest peak is not at the signal period, the lower LS power at the signal period is recorded instead.

A factor to consider for the FAP$_\phi$ component is that starspot evolution may lead to phase inconsistencies across the long-term Evryscope observations, even in real astrophysical signals. Such a situation would decrease the strength of the FAP$_\phi$ component of FAP$_\mathrm{tot}$. If the phase is consistent, however, the detection can be more securely made.

\subsection{Validation with injection and recovery tests}\label{inj_rec_tests}
We then test our results by injecting sinusoids of the same period and amplitude as the signals from each candidate into randomly-selected Evryscope light curves that do not have rotational modulation. For each candidate, 1000 separate injection and recovery tests are performed. In each test, the sinusoidal phase is left free while the amplitude and period are both fixed. Prior to computing the LS periodogram of each injected signal, the light curve is pre-whitened with the same Gaussian 1D kernel and sine fits as described for the original signal to ensure consistency. LS periodograms of the injected signals are computed and the power at the injected period is recorded. Results from the strongest signals are given in Section \ref{FAP_results}.

\begin{figure*}
	\centering
	{
		\includegraphics[trim= 1 1 1 1, width=0.85\textwidth]{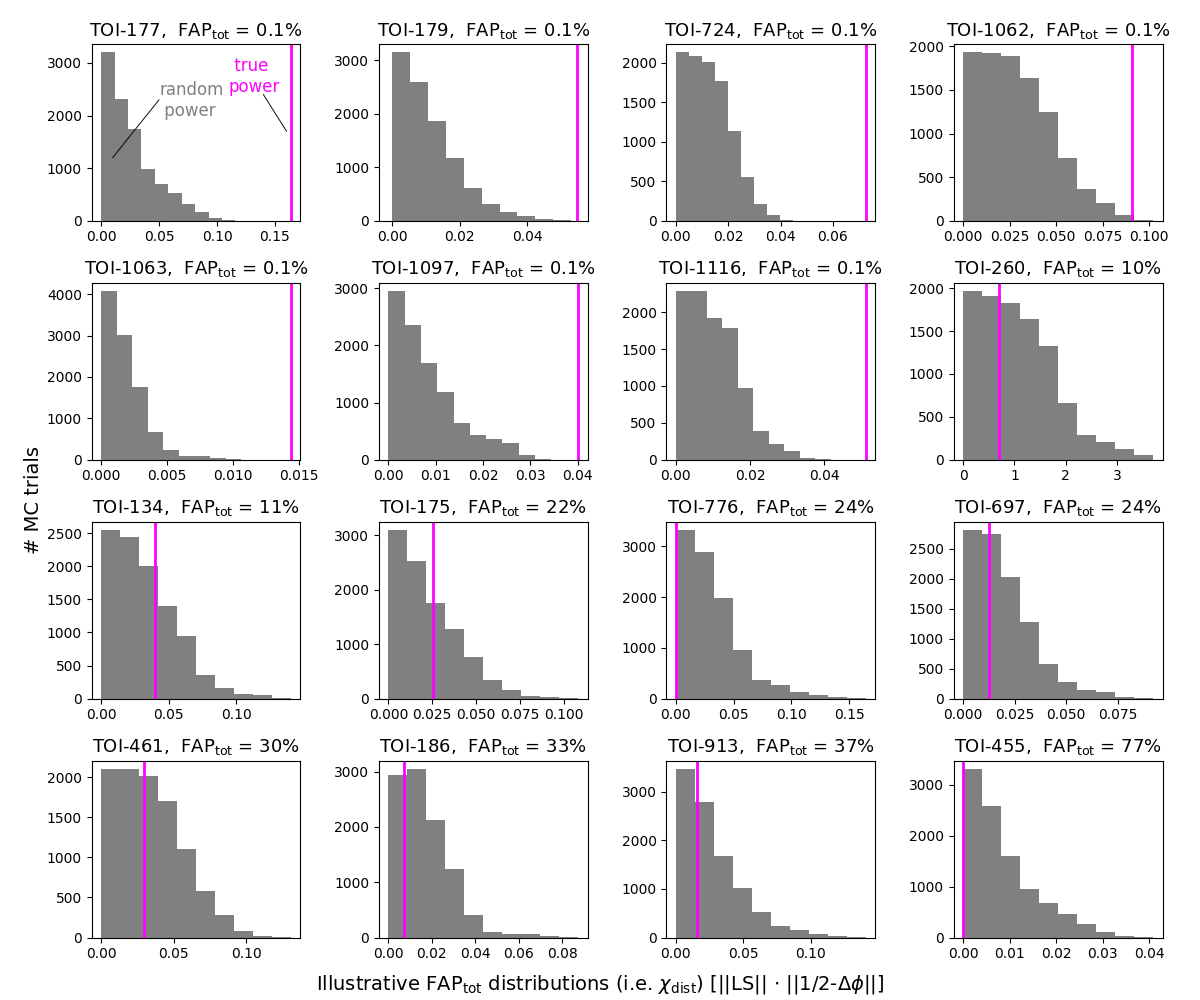}
	}
	\caption{We confirm that FAP$_\mathrm{tot}$ values computed separately from the FAP$_\mathrm{LS}$ and FAP$_\mathrm{\phi}$ distributions qualitatively agree with values from a singular false-alarm distribution made up of random LS$\cdot \phi$ values multiplied together. The resulting $\chi_\mathrm{dist}$ distributions are computed according to Equation \ref{eq:chi_rho_distrib}. Distributions of random LS powers and phase offsets are normalized to their median values and then multiplied together. The resulting distributions are then compared against the observed values, shown in purple. Because larger LS values denote stronger signals and larger $\Delta \phi$ values denote weaker signals, 1/2-$\Delta \phi$ is used instead of $\Delta \phi$ itself. The normalization by the median ensures strong LS signals and strong $\Delta \phi$ signals result in larger LS $\cdot$ (1/2-$\Delta \phi$) values. Values of FAP$_\mathrm{tot}$ reported here and in the main text are computed by multiplying the individual FAP values from the FAP$_\mathrm{LS}$ and FAP$_\mathrm{\phi}$ distributions as verified by the multiplied distributions shown.}
	\label{fig:chiRhoFAP}
\end{figure*}

\begin{figure*}
	\centering
	{
		\includegraphics[trim= 1 1 1 1,clip, width=0.85\textwidth]{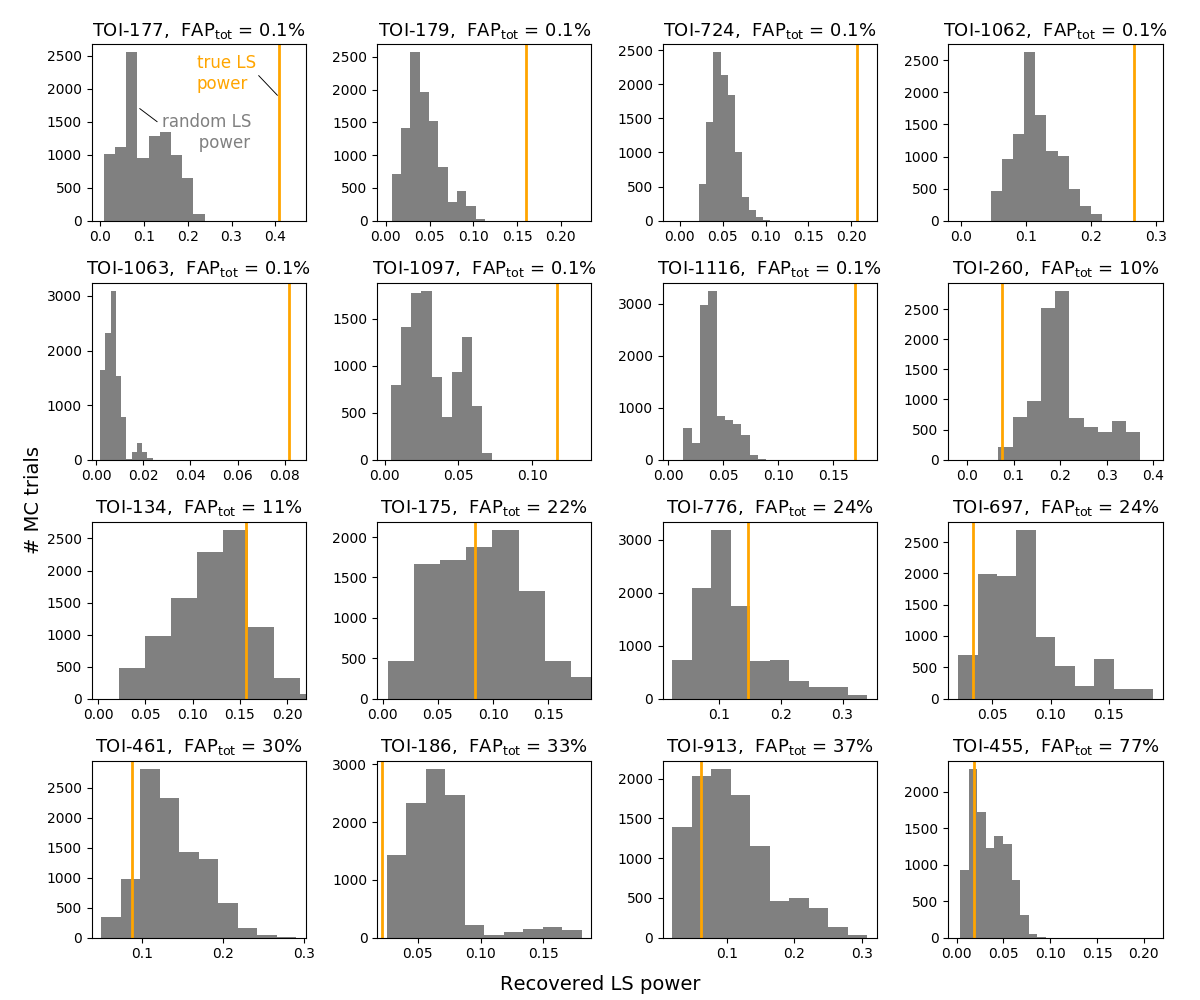}
	}
	\caption{False alarm probabilities from LS periodogram signals (FAP$_\mathrm{LS}$) of the 16 best candidate rotators in our sample. The candidates are ordered by decreasing FAP$_\mathrm{tot}$ instead of decreasing FAP$_\mathrm{LS}$ to account for the contribution of the similarity in sinusoidal phase between Evryscope and TESS. The LS power of each TOI is compared with the distribution of LS power from random light curves across 10,000 MC trials. Random light curves are created by shuffling the magnitudes and times of the target star to preserve the window function then superimposing Evryscope systematics from a random star. The maximum LS peak in the period range ($P_\mathrm{min}$, $P_\mathrm{max}$) is recorded in each trial as described in the text.}
	\label{fig:FAP_results_fig}
\end{figure*}

\begin{figure*}
	\centering
	{
		\includegraphics[trim= 1 1 1 1,clip, width=0.85\textwidth]{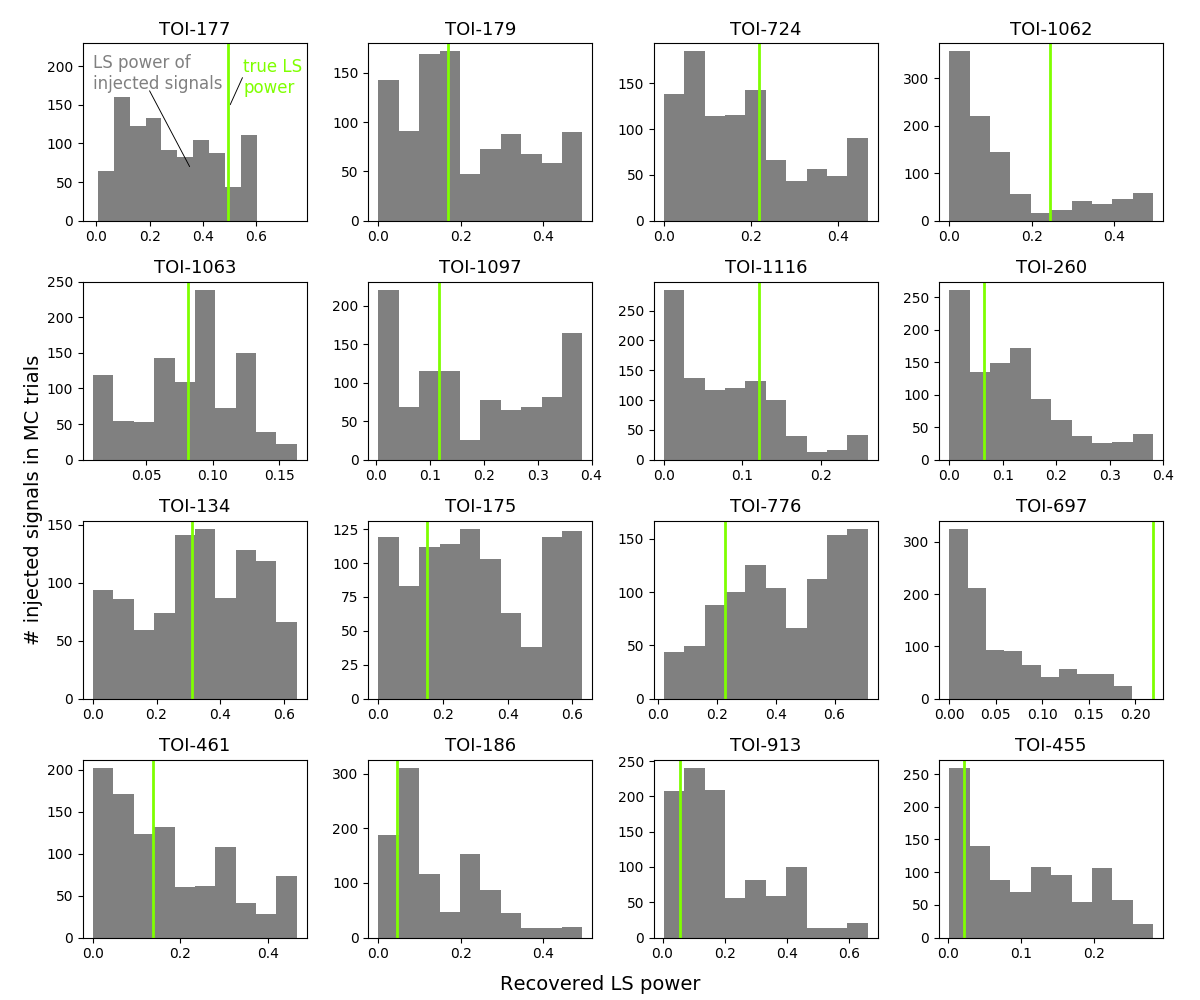}
	}
	\caption{Injection and recovery tests validating that the candidate period for each TOI has a similar LS power as known signals of similar period and amplitude. TOIs are ordered by decreasing FAP$_\mathrm{tot}$ as before. The LS power of the candidate signal is shown in green and compared to the distribution of powers recovered from light curves with injected sinusoids of the same period and amplitude but various phases. The LS power is recorded at the precise period injected in each trial rather than across the whole period range in order to determine if the power we detect is consistent or not with what we would expect from known signals. We do indeed find that the power of candidate signals are consistent with what would be expected. Since we only test the injected period, recovered powers may not always be the dominant peak in the periodogram due to systematics at other periods.}
	\label{fig:inj_rec_fig}
\end{figure*}

\begin{figure*}
	\centering
    \subfigure
	{
		\includegraphics[trim= 0 20 0 10, clip, width=3.0in]{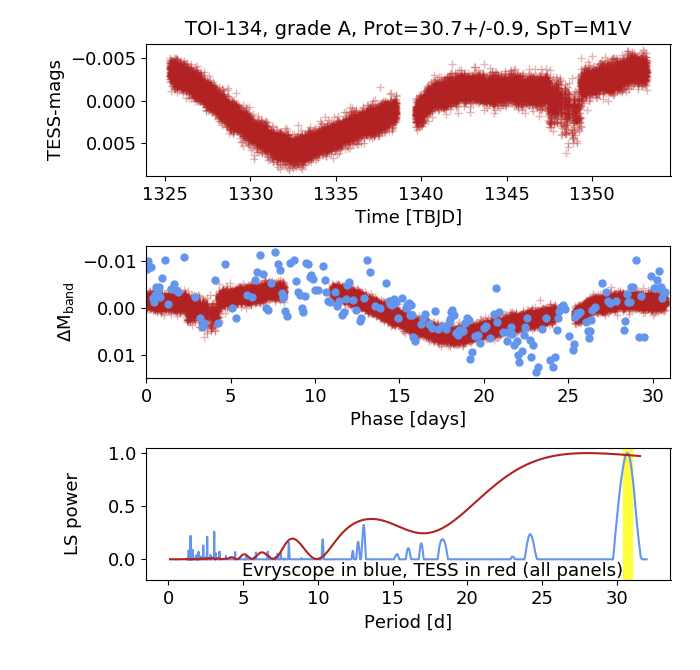}
		\label{fig:TOI_134}
	}
	\vspace{-0.15cm}
	\subfigure
	{
		\includegraphics[trim= 0 20 0 10, clip, width=3.0in]{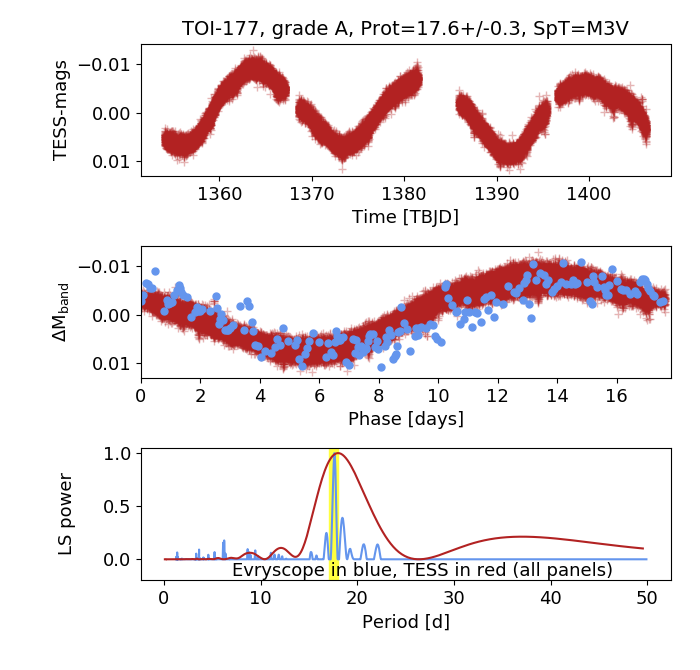}
		\label{fig:TOI_177}
	}
	\vspace{-0.15cm}
	\subfigure
	{
		\includegraphics[trim= 0 20 0 5, clip, width=3.0in]{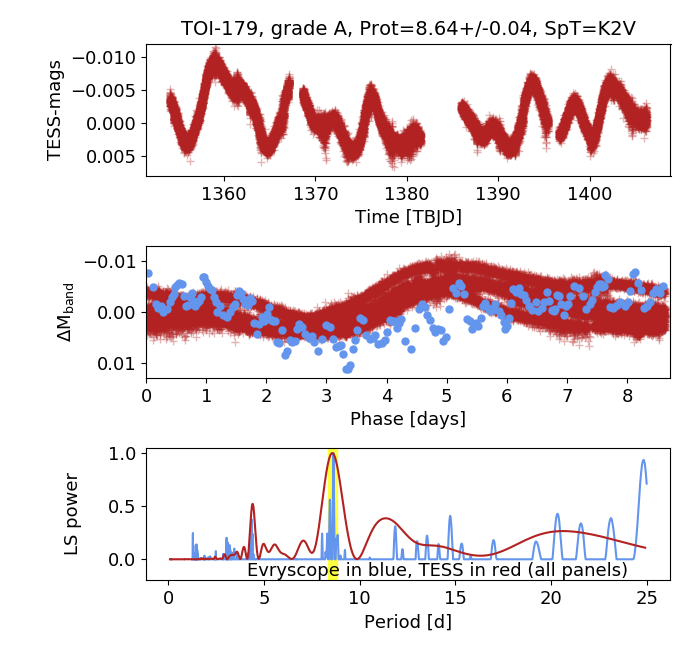}
		\label{fig:TOI_179}
	}
	\vspace{-0.15cm}
	\subfigure
	{
		\includegraphics[trim= 0 20 0 5, clip, width=3.0in]{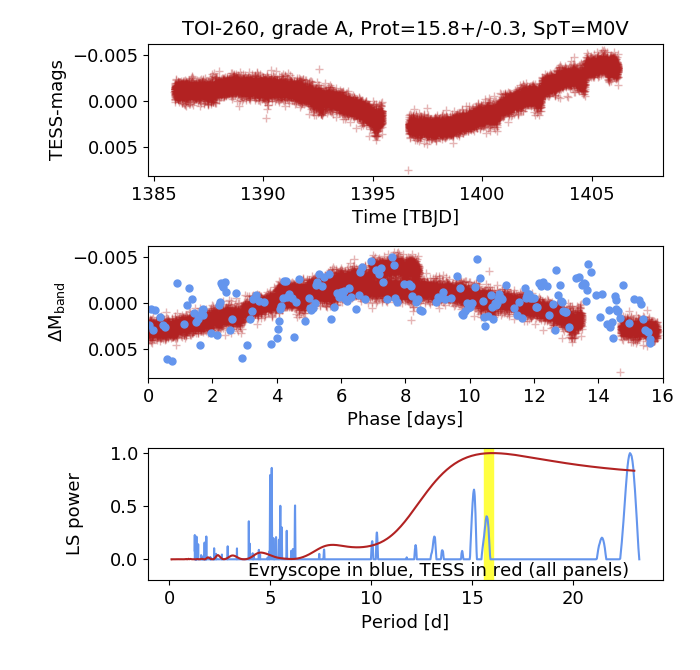}
		\label{fig:TOI_260}
	}
	\vspace{-0.15cm}
	\subfigure
	{
		\includegraphics[trim= 0 20 0 5, clip, width=3.0in]{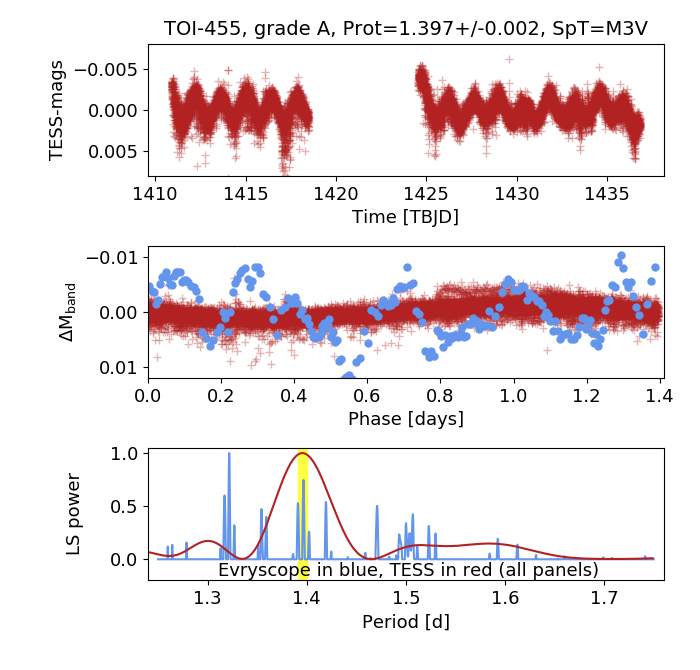}
		\label{fig:TOI_455}
	}
	\vspace{-0.15cm}
	\subfigure
	{
		\includegraphics[trim= 0 20 0 5, clip, width=3.0in]{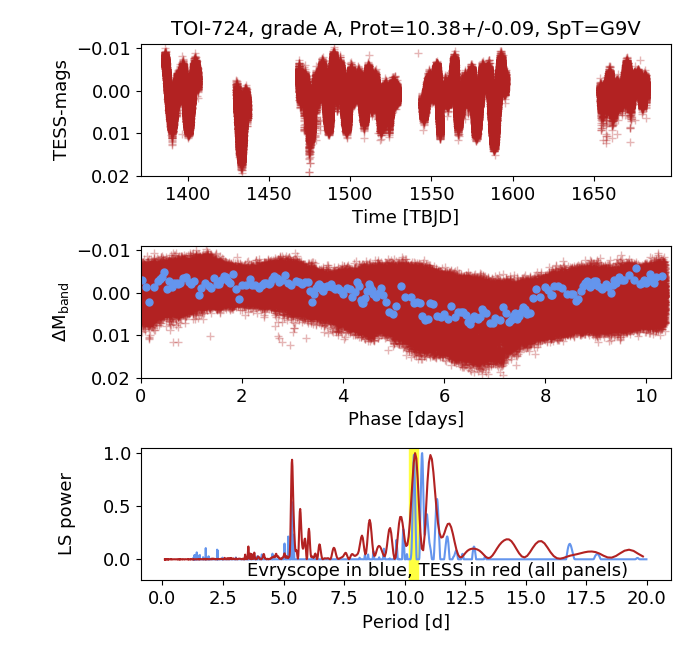}
		\label{fig:TOI_724}
	}
	\vspace{0.01cm}
	\caption{Detection plots for the first 6 out of 10 grade ``A" rotators. Not all grade A rotators are statically confirmed to 3$\sigma$. TOIs 134, 177, 179, 260, 455, and 724 are shown. TOI 455 is only included because of the clarity of the TESS period. The strange behavior of the Evryscope light curve of TOI 455 is most likely pseudo-periodic variation due to instrumental effects rather than the rotation period of another stellar component (from the change in the dilution factor because of Evryscope's smaller pixel scale). In each plot, the TOI number is listed at the top, along with the grade, rotation period, and spectral type (SpT). Top panel: the unfolded TESS light curve. Middle panel: the TESS and Evryscope light curves phase-folded to the detected rotation period, with the Evryscope light curve binned in phase. Bottom panel: Periodograms of the TESS and Evryscope light curves, with rotation period highlighted as a vertical yellow line. As discussed in the main text, if aliasing appears to be splitting power between two Evryscope peaks, both are tested and the one that best matches the phase-folded TESS light curve or has the smallest scatter is selected. This can be seen in TOI 260, for example. In the case of TOI 724, other TESS and Evryscope peaks near the selected signal do not phase-fold as cleanly as the selected peak. This could result from differential rotation and spot evolution across the 2 yr Evryscope light curve, where the upwards error may be under-estimated.}
	\label{fig:first6_gradeA_TOI}
\end{figure*}

\begin{figure*}
	\centering
	\subfigure
	{
		\includegraphics[trim= 0 20 0 5, clip, width=3.1in]{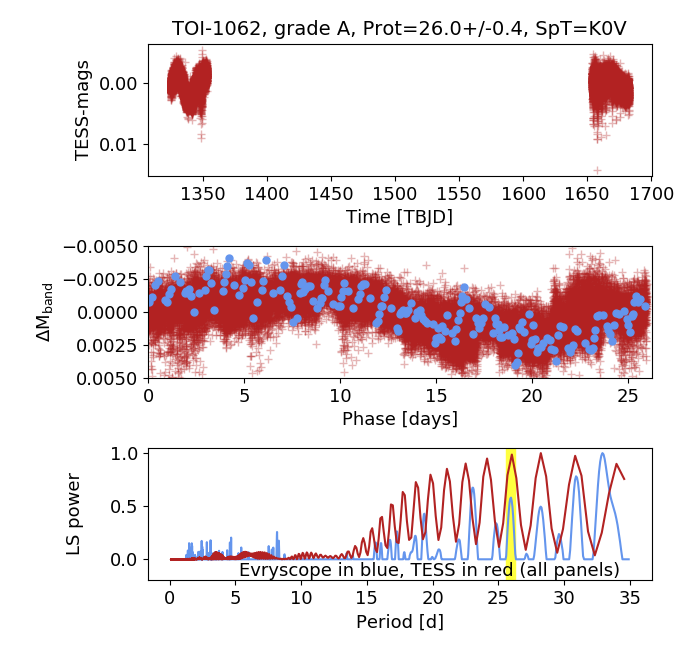}
		\label{fig:TOI_1062}
	}
	\subfigure
	{
		\includegraphics[trim= 0 20 0 5, clip, width=3.1in]{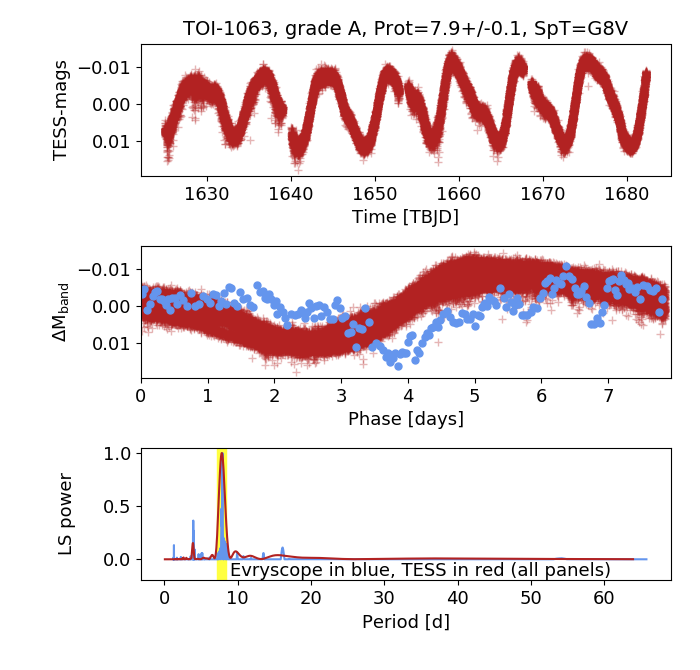}
		\label{fig:TOI_1063}
	}
	\subfigure
	{
		\includegraphics[trim= 0 20 0 5, clip, width=3.1in]{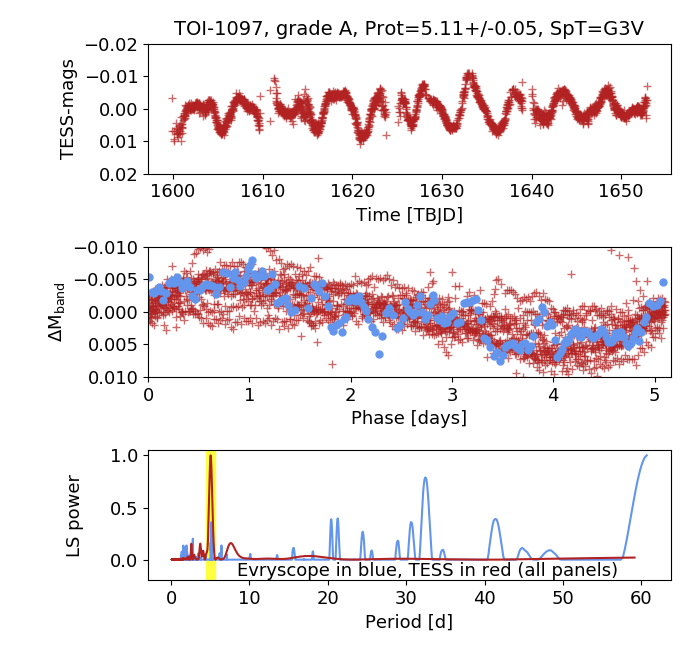}
		\label{fig:TOI_1097}
	}
	\subfigure
	{
		\includegraphics[trim= 0 20 0 5, clip, width=3.1in]{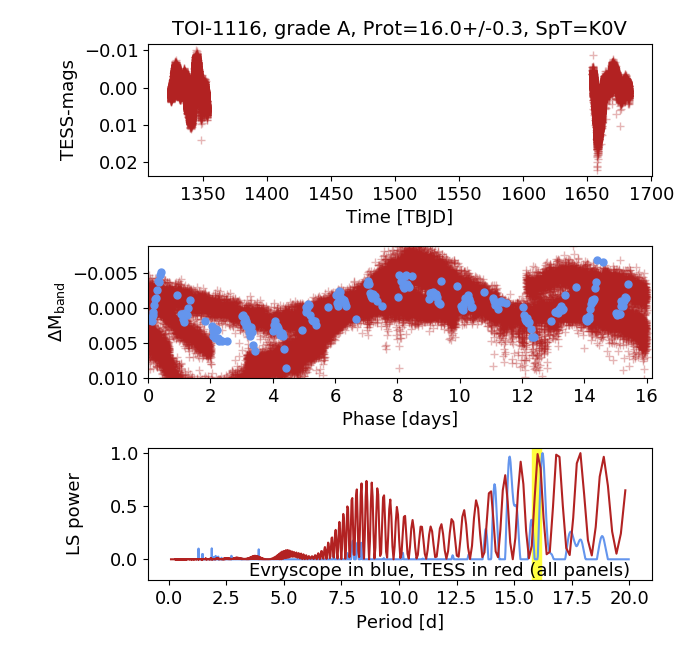}
		\label{fig:TOI_1116}
	}
	\caption{Detection plots for the last 4 out of 10 grade ``A" rotators. Not all grade A rotators are statically confirmed to 3$\sigma$. TOIs 1062, 1063, 1097, and 1116 are shown. In each plot, the TOI number is listed at the top, along with the grade, rotation period, and SpT. Top panel: the unfolded TESS light curve. Middle panel: the TESS and Evryscope light curves phase-folded to the detected rotation period, with the Evryscope light curve binned in phase. Bottom panel: Periodograms of the TESS and Evryscope light curves, with rotation period highlighted as a vertical yellow line. We note that a variable baseline for each periodogram is employed, with the upper limit determined by the length of the unfolded TESS light curve or the ability to clearly see the region around the candidate periods. In the case of TOI 1116, the other Evryscope peak near the selected signal does not successfully phase-fold the TESS and Evryscope light curves, making the error bar based on the FWHM of the selected Evryscope peak alone very plausible.}
	\label{fig:next6_gradeA_TOI}
\end{figure*}

\section{Results}\label{results}

\subsection{Statistical confirmation of grade A and B rotators}\label{FAP_results}
The FAP$_\mathrm{tot}$ of each grade A and B candidate is computed as described in Section \ref{FAP_obj} and tabulated in Table \ref{table:rotation_per_tab}. In Figure \ref{fig:chiRhoFAP}, we illustrate the FAP$_\mathrm{tot}$ values of each rotator via the $\chi_\mathrm{dist}$ distributions described above in \S \ref{FAP_obj}. We also plot the distribution of random LS power and the actual signal power in Figure \ref{fig:FAP_results_fig}, ordered by increasing FAP$_\mathrm{tot}$. The distributions shown in Figure \ref{fig:FAP_results_fig} were used to compute FAP$_\mathrm{LS}$. We do not show plots for FAP$_\phi$ since we compared the phase offsets of the real signals against values from a uniform random phase distribution.

Our FAP analysis finds TOIs 177, 179, 724, 1062, 1063, 1097, and 1116 to be secure $\geq$3$\sigma$ detections. TOIs 134 and 260 are at the 10\% level. TOIs 175, 186, 461, 697, 776, and 913 are more likely to be real than due to chance (FAP$_\mathrm{tot}<$50\%, most 20-30\%) even under the most conservative $\Delta \phi$ limits. The rest cannot be confirmed using the FAP$_\mathrm{tot}$ method. Our method is optimized to give reasonable results in a large sample of rotators but may not be effective for specific low-SNR signals like TOIs 175 and 186 where the period is already known. 3$\sigma$ detections are determined by a FAP$_\mathrm{tot}$ value less than 0.3\% and 2$\sigma$ signals are determined by a FAP$_\mathrm{tot}$ value of less than 5\%, etc. as verified by the distributions of Figure \ref{fig:chiRhoFAP}. We note FAP$_\mathrm{tot}$ is anchored to a traditional $\sigma$-based confidence system through the $\chi_\mathrm{dist}$ distribution. This is because the FAP$_\chi$ false-alarm probabilities are computed in the traditional way from just one input distribution, instead of two input distributions. Most FAP$_\chi$ values qualitatively agree with FAP$_\mathrm{tot}$ values.

The results of the signal injection and recovery tests for each of the top 16 candidate rotators is shown in Figure \ref{fig:inj_rec_fig}. In each case, the real candidate falls in the same LS peak range as the 1000 injected signals of the same period and amplitude. The LS peaks of the strongest candidates at the top of the figure occur more often at the upper end of the distribution of injected signal power, while weaker candidates sometimes occur at the lower end. The type of injection and recovery tests we perform can only statically dis-confirm candidates if they are 3$\sigma$ below the LS peak range. Since this is not the case for our rotators in Figure \ref{fig:inj_rec_fig}, we may only claim they are consistent with the injected signal power.

\subsection{Grade A Rotators}\label{agrade_rotators}
We discovered 10 TOIs with clearly-detected astrophysical variability: these include TOIs 134, 177, 179, 260, 455, 724, 1062, 1063, 1097, and 1116.

\subsubsection{TOIs with multiple period cycles in TESS}\label{multiple_TOIs}
The following TOIs all display multiple observed period cycles in their TESS light curves, making the period determination very straightforward. The Evryscope and TESS rotation period discoveries are shown in detail in Figures \ref{fig:first6_gradeA_TOI} and \ref{fig:next6_gradeA_TOI}. Multiple observed cycles of the period allow us to compute the TESS-only period of most rotators with a GP stellar activity and rotation model from the \texttt{exoplanet} Python package, described in detail in Section \ref{GP_section}. 

\begin{itemize}
  \item TOI 177 (HIP 6365): A nearby (22 pc) M3 dwarf with a candidate 1-2 R$_\oplus$ planet listed on ExoFOP-TESS, TOI 177.01. The star has a $g^{\prime}$ mag of 12.2 and a TESS mag of 9.5. Evryscope observed 24569 epochs over the course of 2.46 yr. We detect a 17.6$\pm$0.3 d stellar rotation period in the combined Evryscope and TESS light curves. Three full periods are present in the TESS light curve, allowing us to compute the TESS-only period to be 17.9$\pm$0.5 with the GP stellar rotation model. An 18 d rotational modulation in WASP data is available for this star on ExoFOP-TESS. Periodicity at 5.21 and 11.17 d is reported to be dubious in the TESS light curves by \citet{Martins2020}.
  \item TOI 179 (HIP 13754): A nearby (39 pc) K2 dwarf with a candidate 2.6 R$_\oplus$ planet listed on ExoFOP-TESS, TOI 179.01. The star is bright in the Evryscope bandpass, with $g^{\prime}$=10.2 and $T$=8.2. Evryscope observed 40471 epochs over the course of 1.96 yr. We detect a 8.64$\pm$0.04 d stellar rotation period in the combined light curves. Three full periods are present in the TESS light curve, allowing us to compute the TESS-only period to be 8.76$\pm$0.1 with the GP stellar rotation model. The TESS light curve demonstrates changes to the amplitude and period. \citet{Martins2020} confirm a 8.489 d period in TESS data alone.
  \item TOI 455 (LTT 1445): A nearby (7 pc) triple system of mid to late M-dwarfs with a confirmed planet, LTT 1445 Ab \citet{Winters2019}. The star is bright in the Evryscope bandpass, with $g^{\prime}$=11.4 and $T$=8.8. Evryscope observed 17051 epochs over the course of 2.08 yr. We detect a 1.397$\pm$0.002 d stellar rotation period in the combined light curves. This is the only Grade ``A" rotator that does not phase-fold to a clear Evryscope sine. 10+ full periods are present in the TESS light curve, allowing us to compute the TESS-only period to be 1.400$\pm$0.004 d with the GP stellar rotation model. \citet{Winters2019} used the same stellar rotation GP \texttt{exoplanet} model that we did and also recovered an identical 1.4 d period. \citet{Martins2020} confirm a 1.393 d period in TESS data alone. The short pseudo-periodic oscillation in the Evryscope light curve is likely systematics rather than the rotation period of another component of the system. It is qualitatively similar to behavior seen in other Evryscope light curves. \citet{Winters2019} note the period likely comes from the active B or C components and not the host star of the planet.
  \item TOI 724 (CD-58 1775): A moderately nearby (95 pc) G9 dwarf with a candidate 2.4 R$_\oplus$ planet listed on ExoFOP-TESS, TOI 724.01. The star is bright in the Evryscope bandpass, with $g^{\prime}$=10.8 and $T$=9.7. Evryscope observed 42084 epochs over the course of 1.99 yr. We detect a 10.38$\pm$0.09 d stellar rotation period in the combined light curves. There are several significant peaks in the TESS and Evryscope periodograms near the 10.38 d signal, but these other signals do not result in as clear of a sinusoidal profile when the light curves are phase-folded. \citet{Martins2020} find a 9.67 d period in TESS data alone.
  \item TOI 1062 (CD-78 83): A moderately nearby (82 pc) K0 dwarf with a candidate 2.3 R$_\oplus$ planet listed on ExoFOP-TESS, TOI 1062.01 (recently confirmed as TOI 1062 b by \citealt{Otegi2021}). The star is bright in the Evryscope bandpass, with $g^{\prime}$=10.6 and $T$=9.5. Evryscope observed 63783 epochs over the course of 1.99 yr. We detect a 26.0$\pm$0.4 d stellar rotation period in the combined light curves. With an amplitude in $g^{\prime}$ of 0.0025, TOI 1062 is the lowest-amplitude long-period signal we confidently detect in Evryscope data. We note the periodicity observed in the TESS light curve may change with time. In addition to the signal strongly detected in Evryscope and TESS at 26 d, there is evidence of a weak signal near 19-21 d that depends primarily on the first sector of data. This signal produces low periodogram power and exhibits high photometric scatter and low amplitude when folded in phase. \citet{Otegi2021} report a $P_\mathrm{Rot}$ of $\sim$22 d from the TESS light curve, $v$ sin $i$, and activity scaling laws.
  \item TOI 1063 (CPD-82 647): A moderately nearby (61 pc) G8/9 dwarf with a candidate 2.3 R$_\oplus$ planet listed on ExoFOP-TESS, TOI 1063.01. The star is bright in the Evryscope bandpass, with $g^{\prime}$=10.2 and $T$=9.1. Evryscope observed 47899 epochs over the course of 1.62 yr. We detect a 7.9$\pm$0.1 d stellar rotation period in the combined light curves. Six full periods are present in the TESS light curve, allowing us to compute the TESS-only period to be 7.88$\pm$0.04 with the GP stellar model.
  \begin{figure*}
	\centering
    \subfigure
	{
		\includegraphics[trim= 0 20 0 10, clip, width=3.1in]{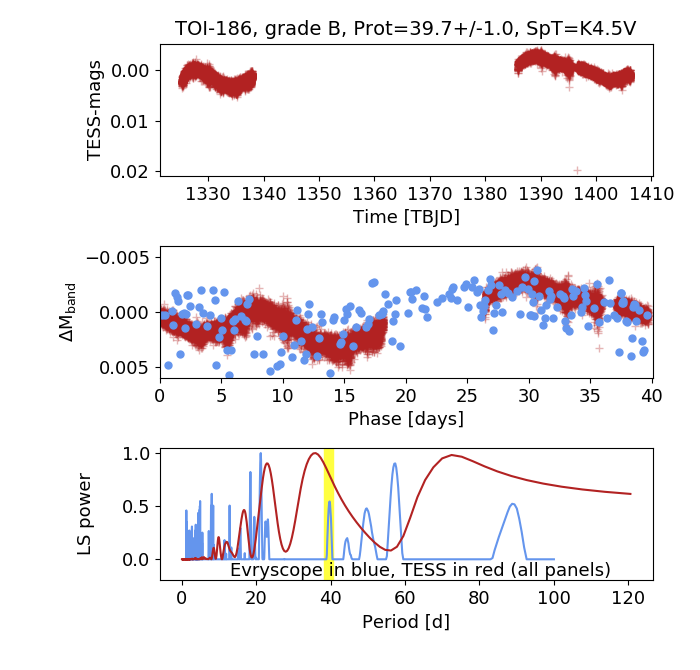}
		\label{fig:TOI_186new}
	}
	\subfigure
	{
		\includegraphics[trim= 0 20 0 5, clip, width=3.1in]{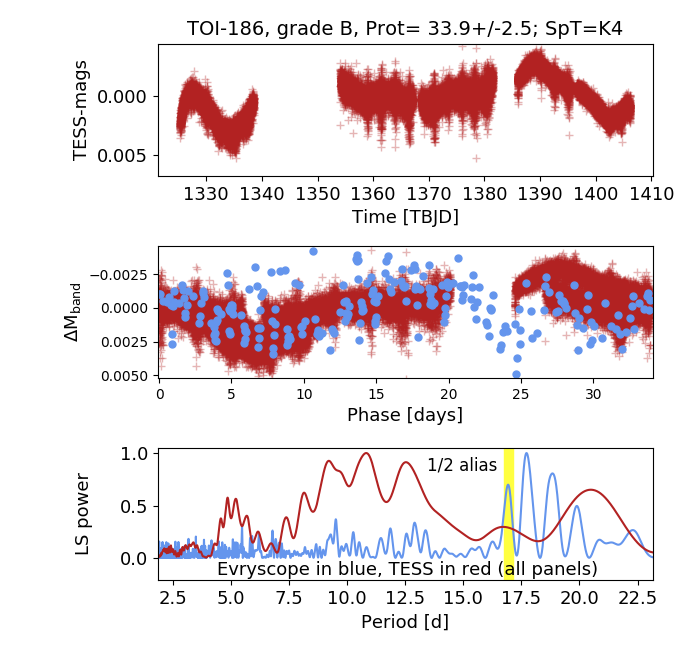}
		\label{fig:TOI_186}
	}
	\caption{Detection plots for the grade ``B" rotator TOI 186. On the left we display the original period detected, and on the right we display the corrected period. The 40 d signal (left) relies on only a subset of the TESS data (noisy data is removed) and may correspond with the observing window; the Evryscope peak is not determinate. On the right, a 33.9 d signal is seen in the TESS and Evryscope light curves. This signal is strongly supported in multiple datasets described in \citet{Gan2021}. Power in the Evryscope LS periodogram appears at the 1/2 alias and not the fundamental period. The 1/2 alias is split into several peaks, with the 16.97 d signal correlating with variability at the same sinusoidal amplitude and phase at twice this period (i.e. 33.9 d) in both the Evryscope and TESS light curves. The TESS periodogram detection at $\sim$33 d is clearer in \citet{Gan2021} who use the Quick Look Pipeline (QLP; \citealt{qlp}) to produce their TESS light curve. Top panels: the unfolded TESS light curve. Middle panels: the TESS and Evryscope light curves phase-folded to the detected rotation period, with the Evryscope light curve binned in phase. Bottom panels: Periodograms of the TESS and Evryscope light curves, with the rotation period signal highlighted as a vertical yellow line. Note the Evryscope periodograms on the left and right are different. This is because the 40 d detection relies on our MP-LS periodogram technique but the 33 d period uses the standard LS periodogram with a different form of pre-whitening (including subtraction of sines at known systematic periods) applied. This is necessary since the MP-LS periodogram heavily suppresses power near 30 d.}
	\label{fig:first_gradeB_TOI}
\end{figure*}
  \item TOI 1097 (HIP 61723): A moderately nearby (80 pc) G3 dwarf with a candidate 2.3 R$_\oplus$ planet listed on ExoFOP-TESS, TOI 1097.01. The star is bright in the Evryscope bandpass, with $g^{\prime}$=10.3 and $T$=8.7. Evryscope observed 42700 epochs over the course of 1.93 yr. We detect a 5.11$\pm$0.05 d stellar rotation period in the combined light curves. Ten full periods are present in the TESS light curve, allowing us to compute the TESS-only period to be 5.10$\pm$0.06 with a GP stellar rotation model.
  \item TOI 1116 (CD-76 73): A moderately nearby (94 pc) K0 dwarf with a candidate 2.4 R$_\oplus$ planet listed on ExoFOP-TESS, TOI 1116.01. The star is bright in the Evryscope bandpass, with $g^{\prime}$=10.6 and $T$=9.5. Evryscope observed 42084 epochs over the course of 1.99 yr. We detect a 16.0$\pm$0.3 d stellar rotation period in the combined light curves. We compute the TESS-only period to be 15.9$\pm$1.1 with a GP stellar activity and rotation model from the \texttt{exoplanet} Python package. The GP fit is shown in the bottom panel of Figure \ref{fig:example_GP_plots}.
\end{itemize}

\subsubsection{TOIs without multiple observed period cycles in TESS}\label{singular_TOIs}
The following TOIs do not have several full period cycles evident in the TESS light curve. The correct period is identified in each case because it is the only LS peak that folds both light curves to the clearest and lowest-scatter sinusoid of the same phase and amplitude.
\begin{itemize}
  \item TOI 134 (L 168-9): A nearby (25 pc) M1 dwarf that hosts a confirmed hot terrestrial planet, L 168-9b \citet{Astudillo-Defru2020arXiv}. The star is bright in the Evryscope bandpass, with $g^{\prime}$=11.8 and $T$=9.2. Evryscope observed 32833 epochs over the course of 1.95 yr. We detect a 30.7$\pm$0.9 d stellar rotation period in the combined light curves. TESS shows good evidence for rotation at periods of $\sim$30 d. Other LS peaks with periods of 13.1, 37.7, and 45.5 d are subtracted from the Evryscope light curve prior to phase-folding at the astrophysical period. The best period in the TESS data alone is 28$\pm$3 d. \citet{Astudillo-Defru2020arXiv} find a rotation period of 29.8$\pm$1.3 d for the star in WASP data. Folding the Evryscope data and the systematics-free first half of the TESS data to a period of 13.1 d also evidences variability of the same phase and amplitude. However, the 30.7 period is preferred since it is evidenced in both the full TESS and Evryscope data.
  \item TOI 260 (HIP 1532): A nearby (20 pc) M0 dwarf with a candidate 1.5 R$_\oplus$ planet listed on ExoFOP-TESS, TOI 260.01. The star is bright in the Evryscope bandpass, with $g^{\prime}$=10.6 and $T$=8.5. Evryscope observed 14982 epochs over the course of 2.11 yr. We detect a 15.8$\pm$0.3 d stellar rotation period in the combined light curves. We choose the second peak in the Evryscope periodogram because it is closer to the highest peak in the TESS periodogram. The best period in TESS is 15.8$\pm$5 d. We subtract sinusoids at likely systematic periods of 5.1, 22.8, and 36.4 d to reduce scatter in the phase-folded Evryscope light curve.
\end{itemize}

\subsection{Grade B Rotators}\label{bgrade_rotators}
\begin{figure*}
	\centering
    \subfigure
	{
		\includegraphics[trim= 0 20 0 10, clip, width=3.1in]{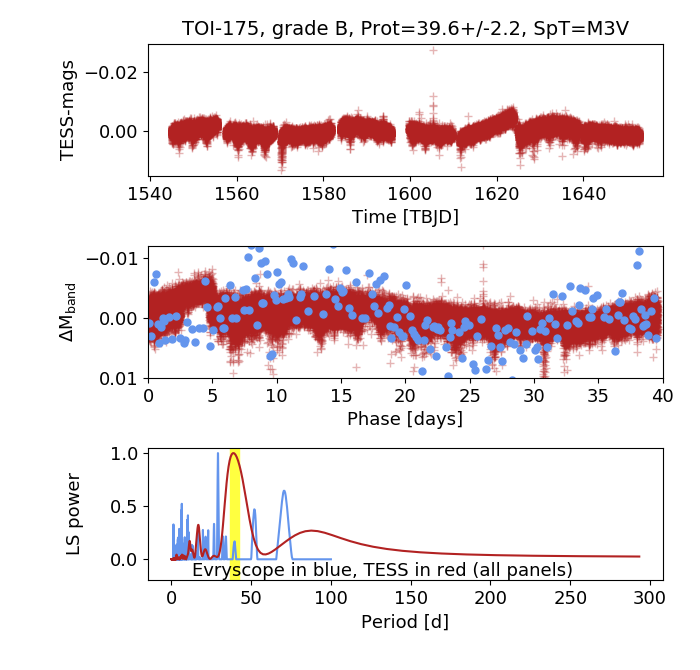}
		\label{fig:TOI_175}
	}
	\subfigure
	{
		\includegraphics[trim= 0 20 0 10, clip, width=3.1in]{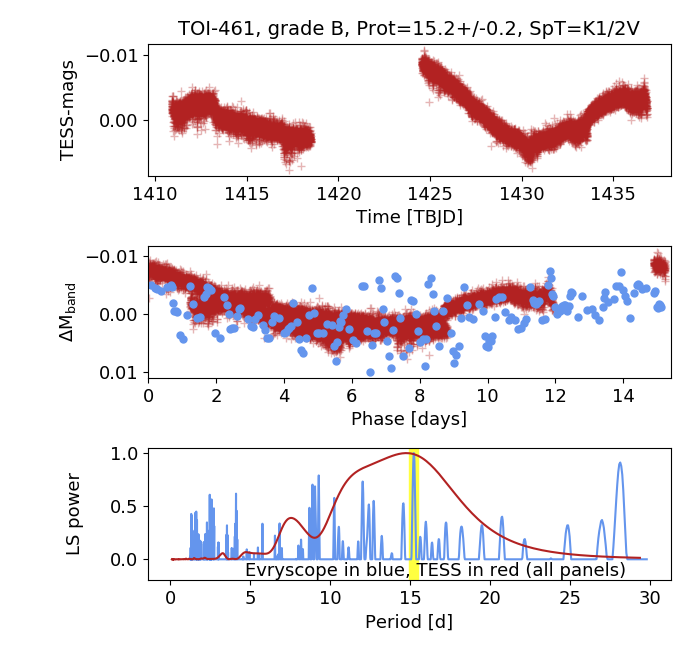}
		\label{fig:TOI_461}
	}
	\subfigure
	{
		\includegraphics[trim= 0 20 0 10, clip, width=3.1in]{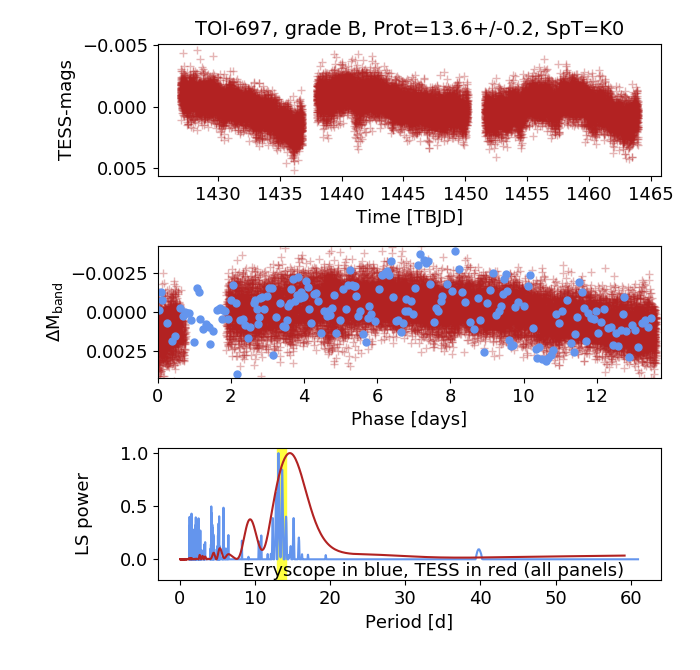}
		\label{fig:TOI_697}
	}
	\subfigure
	{
		\includegraphics[trim= 0 20 0 5, clip, width=3.1in]{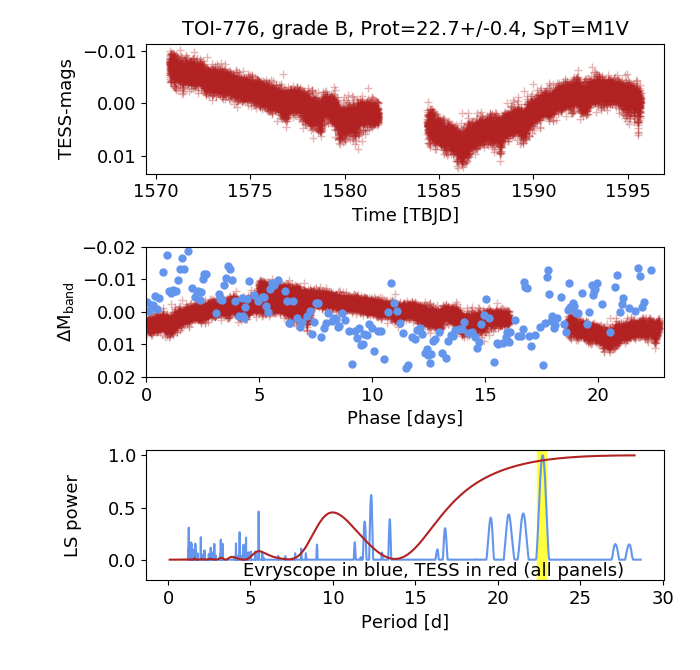}
		\label{fig:TOI_776}
	}
	\subfigure
	{
		\includegraphics[trim= 0 20 0 5, clip, width=3.1in]{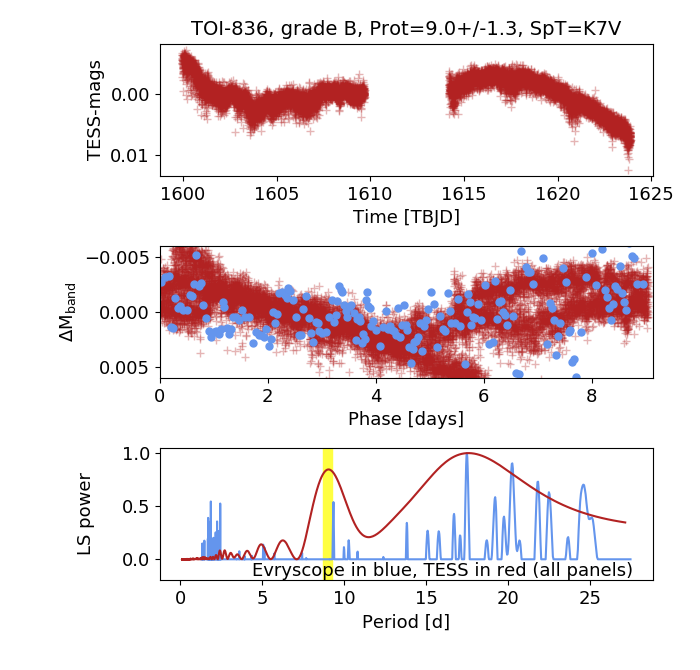}
		\label{fig:TOI_836}
	}
	\subfigure
	{
		\includegraphics[trim= 0 20 0 5, clip, width=3.1in]{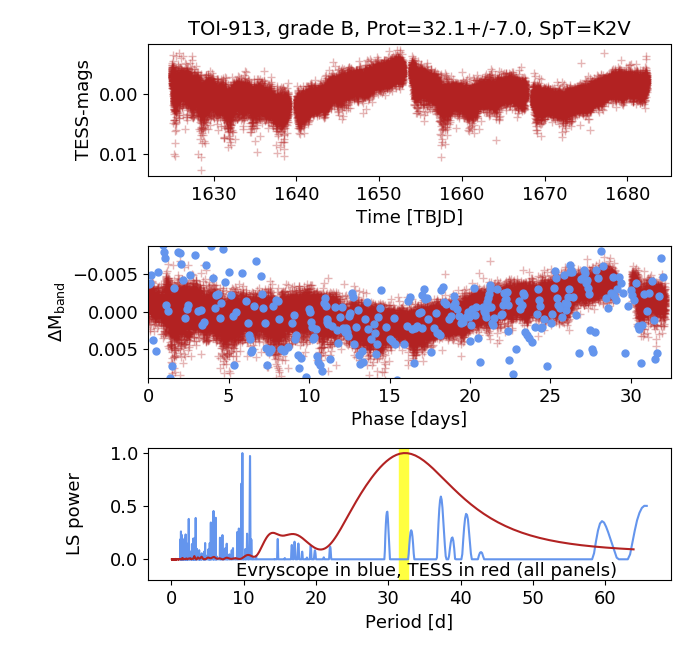}
		\label{fig:TOI_913}
	}
	\caption{Detection plots for 6 out of 7 grade ``B" rotators. TOIs 175, 461, 697, 776, 836, and 913 are shown. In each plot, the TOI number is listed at the top, along with the grade, rotation period, and SpT. Top panel: the unfolded TESS light curve. Middle panel: the TESS and Evryscope light curves phase-folded to the detected rotation period, with the Evryscope light curve binned in phase. Bottom panel: Periodograms of the TESS and Evryscope light curves, with rotation period highlighted as a vertical yellow line. We note that a variable baseline for each periodogram is employed, with the upper limit determined by the length of the unfolded TESS light curve or the ability to clearly see the region around the candidate periods.}
	\label{fig:all6_gradeB_TOI}
\end{figure*}

We discovered 7 rotators classified as grade ``B" that are plausibly astrophysical signals, including TOIs 175, 186, 461, 697, 776, 836, and 913. The Evryscope and TESS rotation period discoveries are shown in detail in Figures \ref{fig:first_gradeB_TOI} and \ref{fig:all6_gradeB_TOI}.

\begin{itemize}
  \item TOI 175 (L 98-59): is a nearby (10.6 pc) M2-3 dwarf with three confirmed terrestrial planets: L 98-59 b is of radius 0.8 R$_\oplus$, L 98-59 c is of radius 1.4 R$_\oplus$, and L 98-59 d is of radius 1.6 R$_\oplus$ \citep{Kostov2019}. The star has a $g^{\prime}$ mag of 12.5 and a TESS mag of 9.4. Evryscope observed 30088 epochs over the course of 1.98 yr. We detect a 39.6$\pm$2.2 d stellar rotation period in the combined light curves. Discontinuities between sectors in the TESS light curve make a clear period identification difficult. The period of TOI 175 is not clearly evident in the data from either survey taken alone, but is the only period that folds both the TESS and Evryscope light curves to a sinusoid with a similar phasing. The Evryscope periodogram has low power at the TESS periodogram peak, although the Evryscope light curve phase-folds to a clear sinusoid at this value. The rotation period of TOI 175 has been previously estimated to be 78$\pm$13 d \citep{Astudillo-Defru2017, Cloutier2019}. Interestingly, \citet{Cloutier2019} note the RVs display a signature of this rotation at half the $\sim$80 d period, leading the authors to use a period of P$_\mathrm{rot} \sim$40 d when measuring planetary masses, very near our period of 39.6 d.
  \item TOI 186 (HD 21749): A nearby (16 pc) K4.5 dwarf with two confirmed planets: HD 21749b is a 2.6 R$_\oplus$ sub-Neptune \citep{Trifonov2019}, and HD 21749c is a 0.9 R$_\oplus$ terrestrial planet \citep{Dragomir2019}. The star is bright in the Evryscope bandpass, with $g^{\prime}$=10.1 and $T$=7.0. Evryscope observed 29059 epochs over the course of 2.46 yr. We originally detected a 39.7$\pm$1.0 d stellar rotation period in the combined light curves. The best period in the SAP-FLUX TESS data is 36$\pm$5 d. Estimates for the rotation period in \citet{Dragomir2019} range from 35 to 39 days, with a best estimate of 38.954$\pm \sim$1 d using data from KELT \citep{Dragomir2019}. The SAP-FLUX light curve demonstrates times of extreme noise, forcing us to remove sections of the light curve. Detrending it with the Quick Look Pipeline (QLP; \citealt{qlp}) produces a much clearer signal at 33.6 d in \citet{Gan2021}. They find the stellar activity indicators support the shorter period and \citet{Martins2020} find a 2$\times$ harmonic of the shorter period. It is possible KELT and Evryscope have similar observing windows for the target, leading to the 40 d period. \citet{Martins2020} find a period of 66.799 d in the TESS light curve, twice that of \citet{Gan2021}. We confirm the $\sim$33 d signal also appears to be present in our data. We pre-whiten the light curve, iteratively removing likely systematics-affected periods including periods at 1 d, 51 d, 90 d, and 365 d. We observe a series of peaks near the 1/2 alias of the $\sim$33 d TESS signal in a LS periodogram of the pre-whitened Evryscope light curve. A peak at 16.97 d near the 1/2 alias is preferred over the other nearby peaks, as phase-folding the TESS and Evryscope light curves to 33.9 d (twice the 16.97 d alias) results in a match to both the amplitude and phase of the sinusoidal variability. The periodogram and phase-folded light curve are shown in Figure \ref{fig:first_gradeB_TOI}. Some caution is warranted in interpreting periodicity in the Evryscope light curve, as HD 21749 has a $g^{\prime}$ magnitude close to the non-linear regime of Evryscope.
  \item TOI 461 (HIP 11865): A nearby (46 pc) K1/2 dwarf with a candidate 2.3 R$_\oplus$ planet listed on ExoFOP-TESS, TOI 461.01. The star is bright in the Evryscope bandpass, with $g^{\prime}$=10.3 and $T$=8.9. Evryscope observed 17541 epochs over the course of 1.96 yr. We detect a 15.2$\pm$0.2 d stellar rotation period in the combined light curves. The TESS period is not well-presented in the TESS light curve but matches the phase and amplitude of the Evryscope detection.
  \item TOI 697 (CD-36 1818): A moderately nearby (94 pc) K0 dwarf with a candidate 2.2 R$_\oplus$ planet listed on ExoFOP-TESS, TOI 697.01. The star is bright in the Evryscope bandpass, with $g^{\prime}$=10.3 and $T$=9.4. Evryscope observed 39767 epochs over the course of 1.97 yr. We detect a 13.6$\pm$0.2 d stellar rotation period in the combined light curves. The TESS period of $\sim$14.638 d is 7\% away from the Evryscope period.
  \item TOI 776 (LP 961-53): A nearby (27 pc) M1 dwarf with two candidate planets listed on ExoFOP-TESS: TOI 776.01 has a radius of 2.2 R$_\oplus$, and TOI 776.02 has a radius of 1.8 R$_\oplus$. The star has a $g^{\prime}$ mag of 12.4 and a TESS mag of 9.7. Evryscope observed 33693 epochs over the course of 2.45 yr. We detect a 22.7$\pm$0.4 d stellar rotation period in the combined Evryscope and TESS light curves. The TESS-only period is poorly constrained at 28$\pm$10 d, a 19\% difference from the Evryscope peak. The TESS and Evryscope light curves both fold to clear sinusoids of comparable amplitudes at the 22.7 d period, but they are offset in sinusoidal phase by 50\%. \citet{Martins2020} record a noisy light curve; \citet{Oelkers2018} record a period of 1.04328 d in KELT data.
  \item TOI 836 (HIP 73427): A nearby (28 pc) K7 dwarf with two candidate planets listed on ExoFOP-TESS: TOI 836.01 has a radius of 2.6 R$_\oplus$, and TOI 836.02 has a radius of 1.8 R$_\oplus$. The star is bright in the Evryscope bandpass, with $g^{\prime}$=10.6 and $T$=8.6. Evryscope observed 23533 epochs over the course of 2.38 yr. We detect a 9.0$\pm$1.3 d stellar rotation period in the combined light curves. Both Evryscope and TESS phase-fold to a sinusoid, and both surveys have power at this period. There is some uncertainty in which periodogram peak is correct, as the two periodograms appear correlated and both show peaks at the same several periods. The 9.04 d signal folded to the simplest sinusoidal shape, so we selected this period.
  \item TOI 913 (CD-80 565): A moderately nearby (65 pc) K2 dwarf with a candidate planet listed on ExoFOP-TESS: TOI 913.01 has a radius of 2.6 R$_\oplus$. The star has a $g^{\prime}$ mag of 11.0 and a TESS mag of 9.6. Evryscope observed 104,018 epochs over the course of 2.97 yr. We detect a 32.1$\pm$7 d stellar rotation period in the combined Evryscope and TESS light curves. Likely systematics-affected periods at $\sim$10 d are subtracted from the Evryscope light curve prior to phase-folding at the astrophysical period.
\end{itemize}

\subsection{Grade U Rotators}\label{ugrade_rotators}
We discovered 12 rotators classified as grade ``U" that are uncertain signals. We do not describe in detail each uncertain signal as we did for grades A and B, but list them here for completeness. These include TOIs 174, 214, 283, 286, 402, 431, 562, 719, 733, 784, and 895, and 1233. TOI 214, 283, 286, 431, 562, and 1233 are more likely to be real than the others. Further details are found in Table \ref{table:rotation_per_tab}. \citet{Martins2020} find a period of 20.5297 d for TOI 562 in TESS data.

\begin{figure*}
	\centering
	{
		\includegraphics[trim= 1 1 1 1,clip, width=\textwidth]{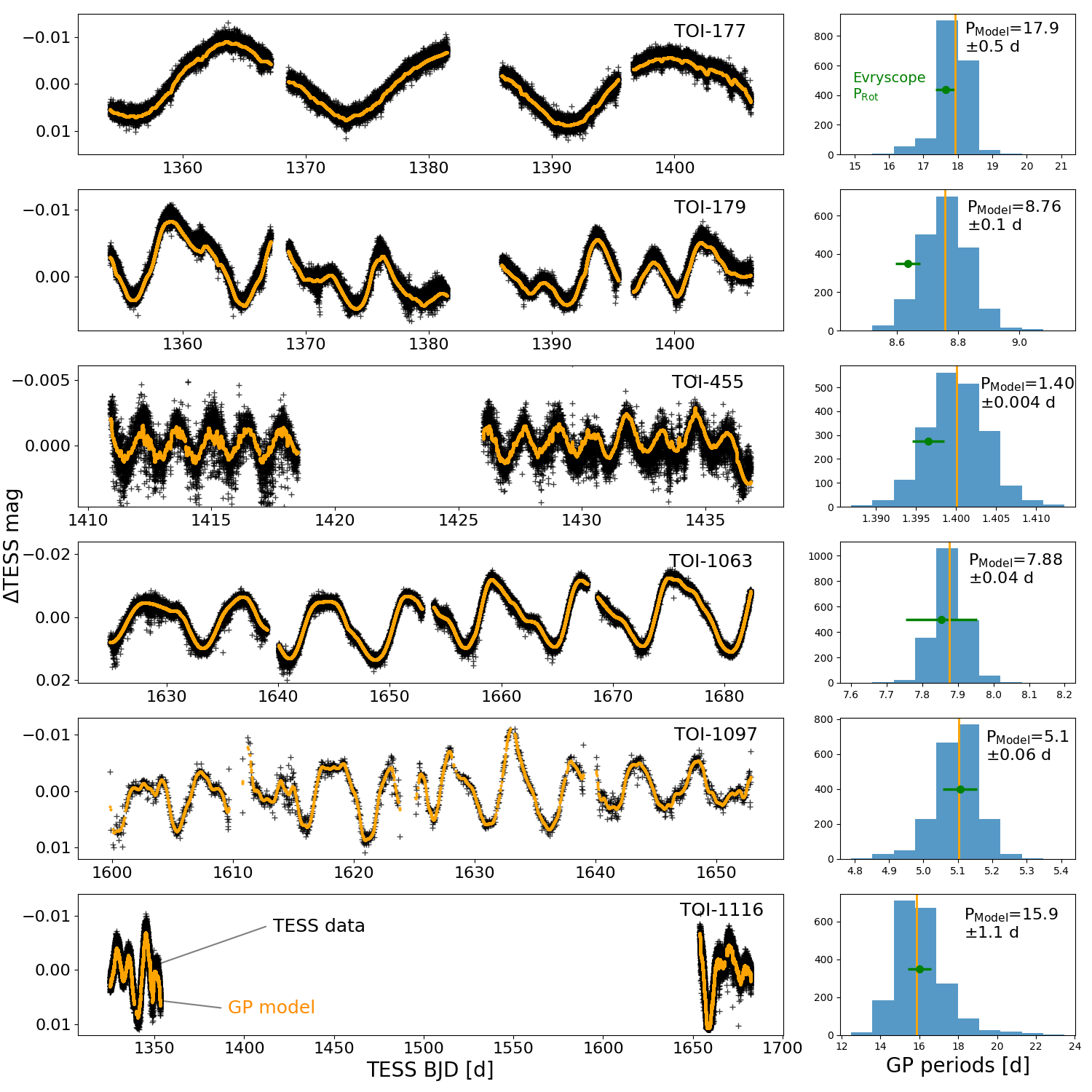}
	}
	\caption{TESS light curves of 6 TOIs with a sufficient number of complete period cycles to constrain the period uncertainty using a GP. The TESS epochs are shown in black and the GP model light curve is shown in orange. To the right of each target star light curve are shown histograms of the periods drawn from the Monte Carlo analysis of the GP model. The median period of the distribution is shown in orange. The Evryscope+TESS periods and uncertainties from the primary analysis of this work are shown for reference as green points with horizontal errorbars. The propagated errorbar of TOI 1116 is enlarged in the plot to make it clearly visible.}
	\label{fig:example_GP_plots}
\end{figure*}

\subsection{Non-detections}\label{ugrade_rotators}
We discovered 6 targets classified as grade ``N" that are non-detections. We do not describe in detail each non-detection as we did for grades A and B, but list them here for completeness. These include TOIs 141, 144, 262, 652, 687, and 1011. Further details are found in Table \ref{table:rotation_per_tab}. \citet{Martins2020} find no period in noisy TESS data for TOI 141, while \citet{Oelkers2018} find a period of 1.12583 d in KELT data.

\section{Verifying stellar rotation with a GP}\label{GP_section}
We use Gaussian Processes (GP) implemented in the Python packages \texttt{exoplanet} to test the robustness of a subset of the periods discovered in Section \ref{agrade_rotators}. A GP is a stochastic model that is composed of a mean function and a co-variance function known as the ``kernel." The GP is parameterized by variables allowing the log-likelihood of the GP to be maximized with respect to those variables; the log-likelihood function is computed using an $N$-dimensional Gaussian \citep{Foreman-Mackey2017}. Quasi-periodic stellar variability due to the rotation of starspots is usually well-described by a GP model with a kernel composed of the sum of two simple harmonic oscillators, e.g. \citet{Foreman-Mackey2020, Winters2019}. We therefore model each light curve in \texttt{exoplanet} with the \texttt{exoplanet.gp.terms.RotationTerm} GP kernel, a sum of two simple harmonic oscillators with the primary oscillation at the rotation period P$_\mathrm{rot}$ and the secondary oscillation at half the period of the primary oscillation. The hyper-parameters of the GP model include the following:
\begin{enumerate}
	\item \texttt{log\_amp}, the log of the amplitude of oscillation.
    \item \texttt{log\_period}, the log of the period of the primary mode of oscillation.
    \item \texttt{log\_Q0}, the quality factor of the secondary mode of oscillation (minus half).
    \item \texttt{log\_deltaQ}, the difference between the quality factors of the primary and secondary modes of oscillation. This value must be positive for the quality factor of the primary mode of oscillation to be of higher quality. 
    \item \texttt{mix}, the amplitude of the secondary mode of oscillation expressed as a fraction of the primary amplitude. 
    \item \texttt{logSw4}, a component of the non-periodic stellar variability.
    \item \texttt{logw0}, another component of the non-periodic stellar variability.
    \item \texttt{logs2}, a description of the stellar jitter, or unaccounted-for white noise.
    \item \texttt{mean}, the average TESS magnitude of the light curve, which should be approximately equal to zero due to pre-processing.
\end{enumerate}

Multi-process Markov Chain Monte Carlo sampling is performed in \texttt{exoplanet} using 2 chains in 28 jobs. We use 1000 draws and 1000 tuning steps for each target star, with a target acceptance rate of 90\%. We explore the variation in the model rotation periods consistent with the light curve. If the posterior distribution of periods has a Gaussian spread, we record the median and 1$\sigma$ values for our period measurement. Period uncertainties do not include occasional values away from the distributions shown in Figure \ref{fig:example_GP_plots}. 

As far as possible, we used the default settings as given in the \texttt{exoplanet} stellar rotation tutorial, ``Gaussian process models for stellar variability \footnote{https://exoplanet-docs.readthedocs.io/en/latest/tutorials/stellar-variability/}." While this usage makes assumptions about the types of noise in the TESS light curves, shape of prior distributions, cyclical decay rate timescales, and number of terms in the rotation kernel, we find in Figure \ref{fig:example_GP_plots} that the default settings do an excellent job in modeling the stellar variability. In these cases, the degree of violations of these assumptions do not appear to dominate the performance of the GP model. Further work exploring the impacts of these assumptions in a larger sample of TESS light curves is encouraged. We refer the reader to the \textsf{exoplanet} \citep{Foreman-Mackey2020} documentation and \citet{Winters2019} for further details.

We initially attempted to measure the periods of both Evryscope and TESS light curves using our GP model. The model performed well on TESS light curves that contained multiple full period cycles, but typically failed to converge for TESS light curves that contained less than 1-2 periods. 

Even the clearest and highest-amplitude rotator in the Evryscope dataset (TOI 177) failed to converge on a period, likely because of the relatively low photometric precision compared to TESS. The rotation GP in \texttt{exoplanet} is probably designed for high-precision \textit{Kepler} and TESS photometry and not for lower precision Evryscope light curves. As described in Section \ref{evr_p_rot_measurements}, the detection of several-mmag rotators in Evryscope light curves requires phase-folding years of data, greatly increasing the photometric precision of the system.

For rotators with several full periods present in the TESS light curve, we compute the GP model period and 1$\sigma$ period uncertainty and compare these values with those from the LS and MP-LS analyses of Section \ref{evr_p_rot_measurements}. These values are reported in Table \ref{table:Bfield_tab} and are also discussed in the target-by-target summaries of ``A"-grade rotators in Section \ref{agrade_rotators} when summarizing the available information for each rotator. We successfully compute GP periods and errors on the TESS light curves of the six A-grade rotators TOIs 177, 179, 455, 1063, 1097, and 1116. The light curves, GP models, and posterior period distributions of these targets are shown in Figure \ref{fig:example_GP_plots}. We also compute a period for the A-grade rotator TOI 260, although the light curve covers only one period. The period uncertainty of TOI 260 is therefore large and does not converge clearly to a Gaussian. Note this is the uncertainty using only TESS information and not Evryscope information, which would make the error smaller as it is in the earlier part of this paper. We also attempted and failed to compute periods for TOI 134 and TOI 461. Of these targets, only TOI 724 had multiple full period cycles present in the TESS light curve but did not properly converge. The period of TOI 724 is 10.4 d and requires multiple sectors to observe several full cycles; the failure may be due to inconsistencies between each sector of data.

\begin{table}
\caption{Stellar rotation GP hyper-parameters}
\begin{tabular}{p{1.4cm} p{1.3cm} p{2.6cm} p{1.8cm}}
\hline
 & & & \\
Hyper-parameters & Prior & Value & Bounds \\
 & & & \\
\hline
 & & & \\
log\_amp & Gaussian & $\mu$=log var($\Delta$T) & $\sigma$=5.0 \\
log\_period & Gaussian & $\mu$=log(P$_{rot}$) & (0.0, log 50.0) \\
log\_Q0 & Gaussian & $\mu$=1.0 & $\sigma$=10.0 \\
log\_deltaQ & Gaussian & $\mu$=2.0 & $>$0, $\sigma$=10.0 \\
mix & Uniform & --- & (0.0-1.0) \\
logSw4 & Gaussian & $\mu$=log var($\Delta$T) & $\sigma$=5.0 \\
logw0 & Gaussian & $\mu$=2$\pi$/10 & $\sigma$=5.0 \\
logs2 & Gaussian & $\mu$=2 log var($\Delta$T$_\mathrm{err}$) & $\sigma$=2.0 \\
mean & Gaussian & $\mu$=0.0 & $\sigma$=10.0 \\
 &  &  & \\
\hline
\end{tabular}
\label{table:Bfield_tab}
{\newline\newline \textbf{Notes.} The list of hyper-parameters governing the GP stellar rotation model. Each hyper-parameter is listed along with the type of prior distribution used, the mean value $\mu$ of that distribution, and the bounds on the distribution (usually a standard deviation limit $\sigma$). Variables include the delta TESS magnitude $\Delta$T, the error in delta TESS magnitude $\Delta$T$_\mathrm{err}$, and the rotation period P$_\mathrm{rot}$ in days. The mix is assigned from a uniform distribution with values between zero and one. \newline}
\end{table}

\section{Summary, Discussion, and Conclusions}\label{discuss_conclude}
As part of the Magellan-TESS survey, we obtain photometric rotation period candidates for 17 unique TOIs that host 1-3 R$_\oplus$ planets and planet candidates. We search the combined Evryscope and TESS light curves of 35 TOIs and 43 planets to find 10 grade ``A" rotators, 7 grade ``B" rotation detections, 12 grade ``U" dubious signals, and 6 grade ``N" non-detections. Only 7 of the grade A periods are confirmed at 3$\sigma$, and no lower grades reach this threshold. We find secure rotation periods that range from 1.4 to $\sim$26 d, and sinusoidal amplitudes of oscillation ranging from 2 to 10 mmag in the Evryscope $g^{\prime}$ bandpass. The sinusoidal amplitudes are similar in the $T$ bandpass.

For rotators with at least 3 full period cycles in the TESS light curves, we use a stellar rotation GP in \texttt{exoplanet} to determine the periods and errors from the TESS light curves alone and compare the results with the full analysis as a spot check on the accuracy of both techniques. We confirm the periods of 6 grade ``A" rotators in this way.

We employ simultaneous light curve and periodogram analyses in the TESS and Evryscope datasets to filter out systematics not common to both surveys. We find this approach in combination with phase-folding 2+ years of observations also lowers the noise floor of the Evryscope periodogram, allowing us to recover sinusoidal amplitudes down to 2 mmag and periods of $\sim$26 d. For the cases where rotation is suggested but not confirmed in the Evryscope and TESS data, we suggest future work to ascertain if the likely effects of activity on the predicted RV jitter can be usefully constrained from timescales of stellar variability not ruled out by our analysis.

It is sometimes difficult to ascertain stellar rotation periods in TESS light curves that span multiple sectors. For example, sector-to-sector discontinuities may be overcome for individual targets using Causal Pixel Modeling (CPM) techniques on light curves (e.g. \citealt{Wang2016}). A CPM application to TESS data to solve the sector continuity problem has recently been created by \citet{Hattori2020}. We find the discontinuity problem can be diminished when light curves from long-term ground based monitoring are available, as TESS systematics and rotation may be more easily separated. We suggest the community employ ground-based light curves in addition to TESS light curves when assessing rotation periods longer than 28 d. Used in conjunction with techniques such as CPM, Evryscope light curves may help to measure periods of a larger sample of TESS rotators \footnote{Until funding is secured to publicly host the Evryscope light curves database, in most cases the Evryscope team is willing to provide individual light curves upon request.}.

\section*{Acknowledgements}\label{acknowledge}
WH acknowledges funding support of this work through NASA NNH18ZDA001N-XRP grant 80NSSC19K0290. HC, NL, JR, and AG acknowledge funding support by the National Science Foundation CAREER grant 1555175, and the Research Corporation Scialog grants 23782 and 23822. HC is supported by the National Science Foundation Graduate Research Fellowship under Grant No. DGE-1144081. JT acknowledges support for this work was provided by NASA through Hubble Fellowship grant HSTHF2-51399.001 awarded by the Space Telescope Science Institute, which is operated by the Association of Universities for Research in Astronomy, Inc., for NASA, under contract NAS5-26555. The Evryscope was constructed under National Science Foundation/ATI grant AST-1407589.
\par This paper includes data collected by the TESS mission. Funding for the TESS mission is provided by the NASA Explorer Program.
\par This research made use of \textsf{exoplanet} \citep{Foreman-Mackey2020} and its dependencies \citep{exoplanet:exoplanet, exoplanet:foremanmackey17, exoplanet:foremanmackey18, exoplanet:pymc3, exoplanet:theano}.
\par This research has made use of the Exoplanet Follow-up Observation Program website, which is operated by the California Institute of Technology, under contract with the National Aeronautics and Space Administration under the Exoplanet Exploration Program.
\par This research has made use of the NASA Exoplanet Archive, which is operated by the California Institute of Technology, under contract with the National Aeronautics and Space Administration under the Exoplanet Exploration Program.
\par This research made use of Astropy,\footnote{http://www.astropy.org} a community-developed core Python package for Astronomy \citep{astropy:2013, astropy:2018}, and the NumPy, SciPy, and Matplotlib Python modules \citep{numpyscipy,Jones2001,matplotlib}.

{\it Facilities:} \facility{CTIO:Evryscope}, \facility{TESS}

\bibliographystyle{apj}
\bibliography{paper_references}

\end{document}